\documentclass[11pt]{article}

\usepackage[utf8]{inputenc}
\usepackage[T1]{fontenc}
\usepackage{graphicx} 
\usepackage{amsmath}  
\usepackage{url}      
\usepackage[shortcuts]{extdash}
\usepackage[margin=1in]{geometry} 
\usepackage{float}    
\usepackage{caption}
\usepackage{booktabs} 
\usepackage{natbib}
\usepackage[colorlinks=true, allcolors=blue]{hyperref}
\usepackage{authblk}

\AtBeginDocument{%
  }

\title{The Persistence of Retracted Papers on Wikipedia}

\author[1]{Haohan Shi\thanks{haohan.shi@u.northwestern.edu}}
\author[2,3]{Yulin Yu\thanks{yulin.yu@kellogg.northwestern.edu, yulinyu@umich.edu}}
\author[3,4,5]{Daniel M. Romero\thanks{drom@umich.edu}}
\author[1,2,6]{Em\H{o}ke-\'{A}gnes Horv\'{a}t\thanks{a-horvat@northwestern.edu}}

\affil[1]{Department of Communication Studies, Northwestern University}
\affil[2]{Northwestern Institute on Complex Systems, Northwestern University}
\affil[3]{School of Information Science, University of Michigan}
\affil[4]{Center for the Study of Complex Systems, University of Michigan}
\affil[5]{Department of Electrical Engineering and Computer Science, University of Michigan}
\affil[6]{Department of Computer Science, Northwestern University}

\date{January 2026}

\begin{document}

\maketitle
\begin{abstract}
Wikipedia serves as a key infrastructure for public access to scientific knowledge, but it faces challenges in maintaining the credibility of cited sources—especially when scientific papers are retracted. This paper investigates how citations to retracted research are handled on English Wikipedia. We construct a novel dataset that integrates Wikipedia revision histories with metadata from Retraction Watch, Crossref, Altmetric, and OpenAlex, identifying 1,181 citations of retracted papers. We find that 71.6\% of the citations were initially problematic and in need of reader-facing repair, defined as those added before the paper's retraction (51.5\%) or introduced afterwards without proper warning (20.1\%). While many are eventually corrected, our analysis reveals that these citations persist for a median of 3.68 years (1,344 days). Through survival analysis, we find that bot-mediated flagging (RetractionBot), open access availability, pre-existing online visibility (e.g., Twitter/X mention counts), and page-level organization (e.g., number of categories on a Wikipedia page) are associated with a higher hazard of correction. Conversely, a paper's established scholarly authority—a higher academic citation count—is associated with a slower time to correction. Our findings highlight how the Wikipedia community supports collaborative maintenance but leaves gaps in citation-level repair. We contribute to CSCW research by advancing our understanding of this sociotechnical vulnerability, which takes the form of a community coordination challenge, and by offering design directions to support citation credibility at scale.

\end{abstract}


\maketitle

\section{Introduction}
Wikipedia is one of the most influential collaborative knowledge infrastructures in the world. Maintained by a decentralized network of human contributors, automated bots, and community governance processes, it produces and maintains a vast, continuously evolving corpus of public knowledge \cite{forte2013defining, halfaker2012bots, geiger2010work}. This content shapes public understanding across scientific, cultural, and political domains. It is increasingly used not only by students and journalists but also by search engines and large language models as a foundational information source \cite{touvron2023llama, brown2020language}. Given its centrality in the modern knowledge ecosystem, the ability of the Wikipedia community to maintain the integrity of its information over time has profound epistemic and societal consequences \cite{fallis2008toward, aragon2021preliminary}.

A growing body of CSCW and HCI research has examined how the Wikipedia community responds to real-world change, particularly in the context of high-attention events such as natural disasters, political controversies, social movements, and celebrities' deaths \cite{roberts2022Wikipedia, kurek2025Wikipedia, keegan2013hot, keegan2012staying, twyman2017blm,keegan2015is}. These studies have shown that Wikipedia can mobilize contributors to rapidly produce and update information in response to new developments, demonstrating the community’s capacity for collaborative sensemaking. However, far less is known about how well Wikipedia's editors and community processes handle low-salience epistemic disruptions like paper retractions—cases where previously accepted knowledge becomes invalidated after its initial citation on the platform. These disruptions are slower, less visible, and often go unnoticed by editors and readers alike. The persistence of such trackable errors highlights a fundamental coordination challenge within large-scale, decentralized communities: how to ensure that critical updates are routed to the specific locations where repair is needed \cite{kittur2008harnessing, forte2012coordination}.

\begin{figure}[t]
\centering
\includegraphics[width=0.8\columnwidth]{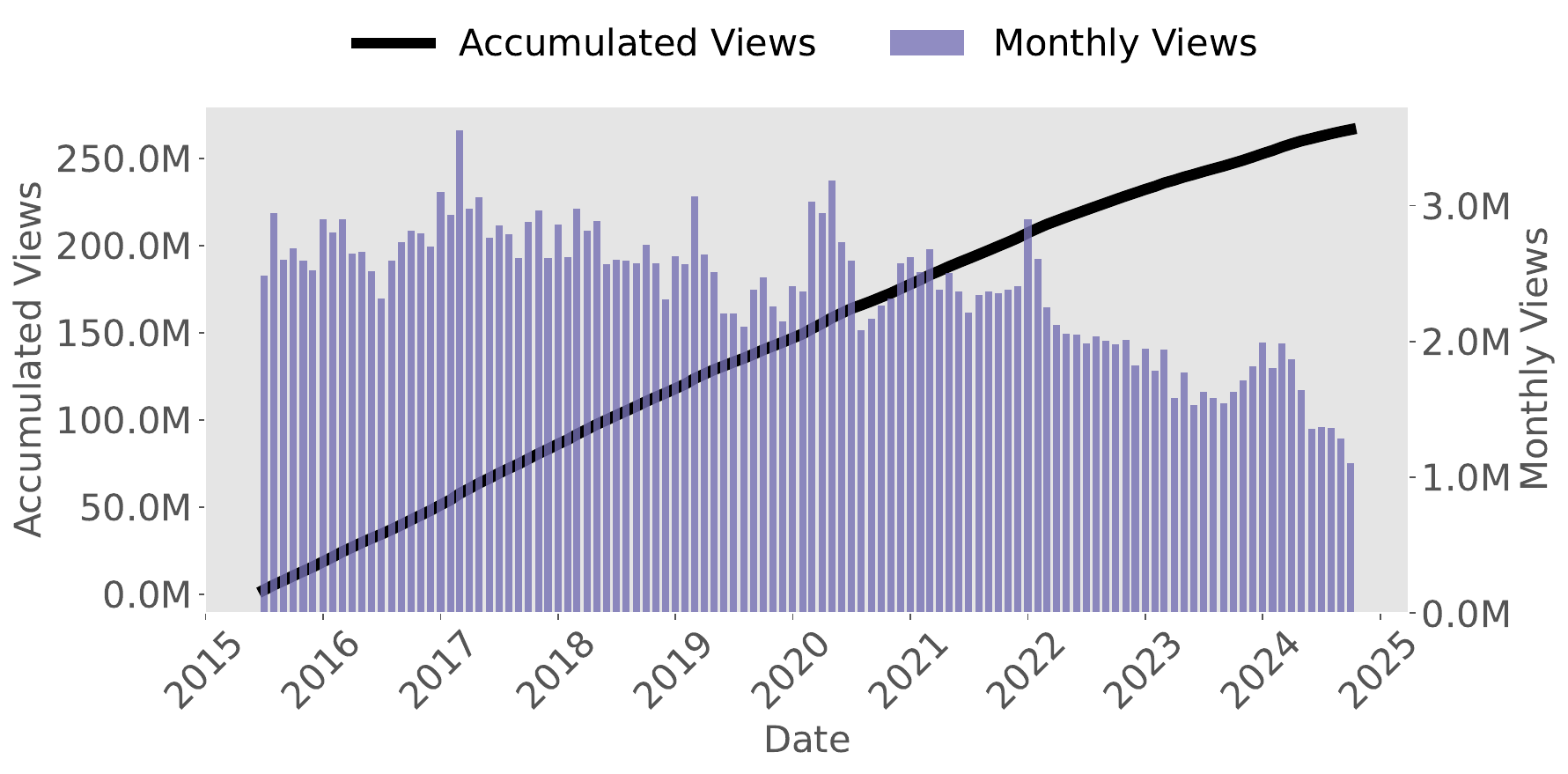} 
\caption{Estimated cumulative and monthly page views of English Wikipedia articles that cite retracted papers without in-text mention of retraction status. Page view data were retrieved using the Wikimedia REST API based on the citation dataset detailed in Section \ref{sec:data_collection}. The figure only shows citations with no accompanying retraction-related keywords (``retract'', ``withdraw'', or ``invalid'') in the main text. This keyword-based detection of uncorrected retraction provides a coarser approximation than the correction categorization used in the main analysis (see Section \ref{sec:data_collection} for details on data collection).}
\label{views}
\end{figure}

Our study focuses on scientific retractions, one of the clearest signals of post-publication epistemic breakdown \cite{jackson201411, fallis2008toward}. Retractions occur when scientific findings are withdrawn due to error, misconduct, or other credibility issues, and they function as formal invalidations of prior knowledge \cite{fang2012misconduct}. When such papers remain cited on Wikipedia without mention of retraction status or removal, they present a critical test of the community’s ability to maintain trustworthy knowledge. From the perspective of readers and other downstream users, this poses serious risks. A person seeking information about a medical treatment, scientific finding, or public health claim may unknowingly encounter content on Wikipedia grounded in invalidated evidence, without any clear warning or correction. Such lapses not only risk misinforming individuals in critical decision-making contexts but also erode trust in Wikipedia as a reliable source of up-to-date knowledge. 

As is illustrated in Figure \ref{views}, English Wikipedia articles containing uncorrected citations to retracted papers have received over 250 million cumulative page views as of 10/24/2024. While not every view results in exposure to the specific citation or misinformation, this figure reflects the substantial reach of flawed content. Moreover, given Wikipedia’s role as a foundational information layer across the web, including in search engine results, educational settings, and machine learning training data, the consequences of uncorrected retractions extend far beyond individual pages and would be difficult to quantify \cite{thelwall2025does}.

Rather than analyzing editor behavior directly, this paper takes a system-level perspective on how the Wikipedia community handles retracted scientific knowledge. Building on CSCW scholarship on collaborative infrastructures, knowledge governance, and epistemic maintenance \cite{ackerman2013sharing, aragon2021preliminary, forte2012coordination}, we examine how citations of retracted papers are distributed across the platform and what observable features are associated with how quickly these citations are corrected. To do so, our study focuses on corrective actions that are directly visible to a typical reader within the main body of an article. We therefore define a ``correction'' narrowly as either the complete removal of the citation or the addition of a visible in-text warning, which includes explicit mentions of the retraction or editorial notes indicating the need for better sources. This definition intentionally excludes automated templates added only to the reference section, which may not be visible to readers who do not consult the bibliography. At the same time, we treat RetractionBot flagging as an observable feature that may be associated with subsequent reader-facing repair. This distinction is important because it recognizes the division of labor within the Wikipedia community. Automated tools are essential for the scalable work of detection and flagging \cite{halfaker2013rise, halfaker2012bots}, providing a vital signal primarily for an editorial audience. However, a complete, reader-facing repair often requires a subsequent act of contextual judgment and editing by humans. Since this human oversight is a key variable and not always guaranteed \cite{kharazian2024governance, baigutanova2023longitudinal}, our study's focus on in-text corrections is designed to measure the success of this complete human-in-the-loop process.

Here, we combine data from Retraction Watch \cite{oransky2012retraction}, Crossref Events \cite{hendricks2020crossref}, Altmetric LLC \cite{altmetric}, OpenAlex \cite{priem2022openalex}, and Wikipedia revision and page view histories to investigate two core aspects of retracted research on Wikipedia:
\begin{itemize}
    \item \textbf{Overview of retracted paper citations on the English Wikipedia.} We explore the distribution of uncorrected citations to retracted papers relative to non-retracted ones and identify the academic domains that disproportionately cite retracted papers. This descriptive overview quantifies the scope of the issue and its potential impact on English Wikipedia users.

    \item \textbf{What factors are associated with the timing of corrective actions for retracted paper citations on English Wikipedia?} This study investigates the factors associated with the timing of corrective actions for retracted paper citations on English Wikipedia. We analyze various features, including a range of features related to the paper itself (e.g., its academic domain, open access status, and academic citation count). Additionally, we examine factors related to the retraction event (e.g. the reasons for retraction, time from publication to retraction, and editorial events before retraction). Finally, we analyze Wikipedia-specific factors such as RetractionBot flagging, page reference count, media count, categories count, and the level of editorial activity (e.g., the number of unique editors). These factors provide insight into the challenges of identifying and addressing problematic citations.
\end{itemize}

Our findings show that Wikipedia’s responsiveness to citations of retracted papers is inconsistent and often slow. While some citations are corrected quickly, many persist uncorrected for years. Among the 1,181 citation pairs analyzed, we found 71.6\% represented problematic cases requiring repair: specifically, 51.5\% were added before the retraction, and 20.1\% were introduced afterwards without in-text warnings. The remaining 28.4\% were classified as non-problematic because they explicitly noted the retraction status in-text at the time of citation. While many of these problematic citations are eventually corrected, our analysis reveals that they persist for a median of over 3.68 years (1,344 days). In the extended Cox model, the correction hazards are higher for citations that are flagged by RetractionBot and for citations that were framed as controversial at the time of addition. Higher correction hazards are also associated with open access papers and with greater Twitter/X attention prior to retraction, although the Twitter/X association is time-dependent and attenuates over time. Page-level structural integration is also associated with faster correction: pages with more categories show higher correction hazards. Conversely, a higher academic citation count is associated with a lower correction hazard, which may reflect the difficulty of challenging sources perceived as highly authoritative. We do not observe a stable average difference by retraction reason, though the association appears to vary over time. Together, these findings indicate that visibility cues and page organization are associated with the timing of citation-level repair on Wikipedia.

Our paper contributes to CSCW research on collaborative knowledge infrastructures in four ways. First, it shifts the focus from Wikipedia’s well-documented responsiveness to high-salience events to its underexamined handling of low-visibility epistemic breakdowns, such as retracted scientific papers. Second, it provides a large-scale empirical analysis of how retracted citations persist—or are corrected—across academic domains, revealing how features such as social media signals, open access status, automated tools (e.g., RetractionBot), and page-level organization (e.g., number of categories) are associated with repair outcomes. Third, it highlights a critical division of labor between reference-level flagging and substantive in-text repair. While our results show that automated interventions (e.g., RetractionBot) significantly accelerate the correction process, they primarily serve as technical signals that catalyze human editorial work rather than replace it. This indicates a broader design gap between machine-generated alerts and editorial workflows: while machine-generated alerts are effective at surfacing issues, the contextual reader-facing correction remains a distinctly human task. Finally, it provides actionable opportunities for design, suggesting how Wikipedia’s tools, workflows, or interfaces might better support editors in identifying and repairing citation-level credibility failures.

\section{Related Work}

\subsection{Wikipedia as an Epistemic Infrastructure}

Wikipedia has served as a prime example in CSCW and HCI research for studying large-scale, decentralized collaboration. As a social computing system, it depends on the interplay of human contributors, automated bots, and governance policies to produce and maintain public knowledge \cite{forte2013defining, halfaker2012bots, geiger2010work}. This sociotechnical infrastructure supports a complex ecosystem of roles, routines, and technical tools that enable collaborative knowledge production \cite{ackerman2013sharing, forte2012coordination}.

Prior work has documented how Wikipedia adapts to rapidly changing external events. In crisis contexts such as natural disasters or breaking news, editors engage in high-tempo coordination, rapidly creating and revising articles to reflect emerging realities \cite{keegan2013hot, keegan2012staying}. Studies of social movement coverage, such as the documentation of Black Lives Matter, extend this perspective to the formation of collective memory in contentious and evolving domains \cite{twyman2017blm}. Keegan et al. \cite{keegan2015is} show how editors update Wikipedia biographies after someone dies, working together to correct facts while also honoring the person’s memory. Recent research on the Russo-Ukrainian War extends this understanding to geopolitical conflicts, highlighting how editors mobilize to manage disruptive editing and maintain information integrity amidst information warfare \cite{roberts2022Wikipedia, kurek2025Wikipedia}.  Together, these studies underscore Wikipedia’s responsiveness, focusing on high-salience events that prompt rapid editorial attention.

Yet recent work highlights Wikipedia’s vulnerabilities in less visible contexts. Kharazian et al. \cite{kharazian2024governance} show how, in smaller language editions of Wikipedia, editorial control can become concentrated in the hands of a few users, allowing misinformation to persist. Similarly, Aragón et al. \cite{aragon2021preliminary} identify risks to knowledge integrity that can arise even without deliberate manipulation. Together, these studies show that Wikipedia’s openness—while foundational to its success—also means that no individual contributor is accountable for the accuracy of any specific claim. When responsibility is diffuse and editorial attention is uneven, epistemic breakdowns can persist without being noticed or repaired. Moreover, Wikipedia’s information credibility can face intentional challenges. Kumar et al. \cite{kumar2016disinformation} studied the creation and detection of Wikipedia hoaxes, fabricated articles designed to deceive readers. Our study contributes to the literature by examining how Wikipedia handles a particular form of non-adversarial epistemic disruption: the citation of scientific articles that have been formally retracted. Failure to maintain the credibility of cited scientific sources leads to a non-negligible breakdown in information quality that is impactful regardless of intention.

Prior work has provided important insights into Wikipedia’s collaborative maintenance processes at the article level. However, fewer studies have examined how the platform manages the credibility of cited knowledge when it becomes formally invalidated. Rather than tracing editorial workflows directly, we analyze the distribution of these citations across Wikipedia and identify factors associated with whether they are eventually corrected or persist unaddressed. This approach highlights how certain infrastructural and visibility features—such as automated flagging, attention signals, and page organization—are associated with whether citations are corrected or persist unaddressed.

\subsection{Scientific Knowledge and Citations on Wikipedia}
Scientific knowledge plays a central role in Wikipedia’s coverage of specialized topics, and recent scholarship has explored how scientific sources are cited, structured, and maintained. Prior work has mapped the landscape of scientific citations on Wikipedia, revealing both the breadth of disciplines represented and the dominant role of biomedical research and papers from high-impact journals. Singh, West, and Colavizza \cite{singh2021Wikipedia} quantified this trend, finding that the life sciences and biomedicine dominate the cited literature, with journals such as Nature and Science among the most frequently referenced. Arroyo-Machado et al. \cite{arroyo2020science} similarly found that multidisciplinary and biomedical journals dominate Wikipedia’s scientific references. Nicholson et al. \cite{nicholson2021measuring} assessed citation quality, showing that Wikipedia tends to cite reputable journals, but still includes a notable fraction of sources with limited academic validation. Other studies have documented representational biases, such as the under-citation of women and non-Anglophone authors \cite{zheng2023gender}.

Complementing this citation-level research, scholars have also examined how Wikipedia’s article quality ratings relate to citation behavior. Arroyo-Machado et al. \cite{arroyo2022wikinformetrics} found that articles rated as higher quality by the Wikipedia community tend to include more academic references, attract more edits, and receive higher page views—but that these associations vary depending on the topical domain. For example, the authors found that a high number of academic references strongly correlates with scholarly-oriented articles in fields like medicine and history. In contrast, high page views are more typical of articles about popular culture, and high talk page activity is associated with controversial subjects. More recently, Baigutanova et al. \cite{baigutanova2023longitudinal} conducted a large-scale longitudinal study of Wikipedia’s reference quality, finding that sourcing practices have generally improved over time, particularly with increasing reliance on peer-reviewed journals and scientific literature. This trend toward higher-quality sourcing builds on earlier observations about Wikipedia’s increasing preference for high-impact journals such as Nature and Science \cite{nielsen2007scientific, wedemeyer2008quality}. Nevertheless, concerns persist about the reliability and maintenance of references, particularly in articles that receive limited editorial attention.

While prior research has primarily focused on the inclusion and quality of citations, relatively little is known about how Wikipedia handles erroneous scientific citations, particularly those referencing retracted research. The study most closely aligned with ours is by Nicholson et al. \cite{nicholson2021measuring}, who identified 118 retracted articles cited across 297 Wikipedia pages. In a manual sample of 50 citations, they found that about half failed to acknowledge the retraction, and even when retractions were mentioned, they were typically used illustratively rather than corrected or flagged clearly. Additionally, their study was based on a small sample. It did not examine what factors are associated with citation correction, thus creating an opportunity for a large-scale characterization of retraction persistence patterns across Wikipedia.

We build on this line of work by identifying retracted papers cited on Wikipedia and analyzing how those citations are handled over time through article revisions. Our study offers a large-scale account of how retracted citations are noted, ignored, or removed. We also model the factors that are associated with these outcomes. In doing so, we extend prior research on citation practices by shifting attention from how scientific citations are sourced and evaluated to how they become invalid and are subsequently addressed. Our work also contributes to CSCW scholarship on collaborative maintenance and infrastructure resilience by examining how Wikipedia editors manage epistemic failures at the citation level.

\subsection{Retractions, Online Attention, and Challenges of Correction}
Retractions are intended to correct the scientific record, but their corrective effects are often limited once flawed research has circulated online. Prior work shows that retracted papers frequently receive more online attention compared to their non-retracted counterparts, particularly on social media, news platforms, and in knowledge repositories \cite{peng2022dynamics}. This attention is often exhausted before the retraction is issued, and, in some cases, the retraction itself may unintentionally amplify visibility. While increased attention can surface public criticism—especially on platforms like Twitter/X \cite{peng2022dynamics, amiri2024tweeters}—only a small and selective group of users tends to engage in such critical discussions. Moreover, retracted papers have also been shown to disseminate false information, particularly around sensitive topics such as vaccination, thereby undermining public trust in science \cite{abhari2024they}. Emerging work has explored whether language models can help flag problematic papers based on these online signals, with promising results using Twitter/X data \cite{zheng2025can}.

While these studies illuminate how retractions garner online visibility, they primarily focus on digital and social media. Much less is known about how collaborative knowledge systems like Wikipedia respond to retracted science. We address this gap by examining retraction correction as a form of sociotechnical repair, and our work contributes to CSCW by extending its focus on knowledge governance, the breakdown of citation credibility, and maintenance in decentralized systems.

\section{Empirical Setup}
\subsection{Data Collection}
\label{sec:data_collection}
\begin{figure}[t]
\centering
\includegraphics[width=1\columnwidth]{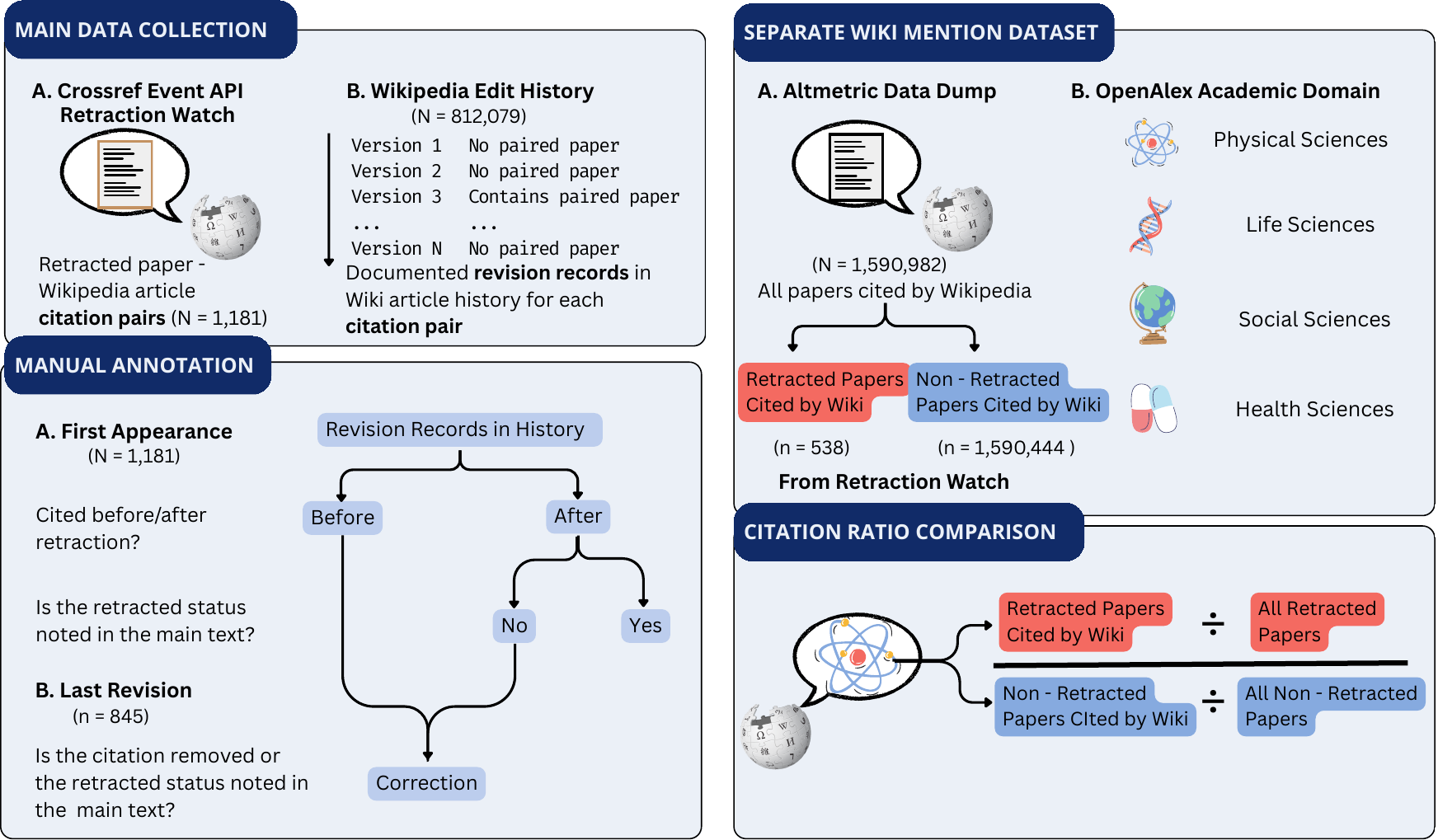} 
\caption{Overview of the data collection and annotation process. The left panel shows the construction of the main dataset, including the extraction of 1,181 retracted paper–Wikipedia citation pairs from Crossref Event Data and corresponding revision histories from Wikipedia. One of the authors manually annotated whether each citation was cited with its retraction status noted at its first appearance and corrected in the latest revision (removal or retraction status noted on the page). The right panel displays the separate dataset of all Wikipedia-cited papers (N = 1,590,982), which is used to calculate citation ratios across academic domains. Retraction status is linked using Retraction Watch, and domains are categorized via OpenAlex.}
\label{collection}
\end{figure}

We integrated data from Retraction Watch \cite{oransky2012retraction}, Crossref Events \cite{hendricks2020crossref}, Altmetric LLC \cite{altmetric}, OpenAlex \cite{priem2022openalex}. We also used the Wikimedia REST API to retrieve monthly page view statistics for articles citing retracted papers, and the MediaWiki Action API to analyze their full revision histories. This integration enables us to systematically examine how retracted research is cited and handled on Wikipedia, allowing us to investigate further how features of papers, Wikipedia pages, online mentions, and temporal trends (e.g., whether citations occurred before or after retraction, and the time between publication and retraction) are associated with the correction of retracted paper citations. 

The Retraction Watch database is the most comprehensive and widely adopted resource on retracted articles \cite{brainard2018rethinking}, indexing all retracted Digital Object Identifiers (DOIs) and their metadata, such as retraction date, publication date, and retraction reason. In this study, we used Retraction Watch to obtain a list of retracted papers and their associated metadata, including article titles, retraction dates, publication dates, and retraction reasons. To ensure our analysis focused exclusively on retracted papers, we excluded the ones with expressions of concern or corrections. Expressions of concern serve as warnings while investigations are ongoing, and corrections address minor issues without invalidating the entire study. Retractions, by contrast, formally withdraw a paper from the academic record. 

We used the Crossref Event API to collect all Wikipedia pages that have cited the retracted DOIs. Crossref periodically crawls Wikipedia and logs pages that contain outbound links to Crossref DOIs, along with detection timestamps and metadata \cite{hendricks2020crossref}. The Crossref Event API is the most reliable in indexing Wikipedia citation data \cite{ortega2018reliability}, and only indexes papers with valid DOIs. 833 retracted papers were cited on 900 Wikipedia pages, forming 1,181 retracted paper--Wikipedia page citation pairs. To accurately date when the retracted DOI was introduced to the Wikipedia pages and when the retracted DOI was properly corrected on Wikipedia (complete removal or adding an in-text warning), we used the MediaWiki Action API to retrieve and analyse all revisions ($N=824,079$) of these pages. This allowed us to perform a detailed temporal analysis of how retracted papers were cited and subsequently addressed on Wikipedia. To ensure our analysis was relevant to public-facing content, we excluded pages from non-article namespaces such as Talk, Sandbox, Draft, and User. 

We constructed a dataset separate from the one used in our main analysis to compare the academic domain distribution of retracted and non-retracted papers cited on Wikipedia. For consistency, we identified 538 retracted papers using the latest Altmetric data dump (current as of August 14, 2023), rather than relying on our original Wikipedia retracted citations dataset. This allowed us to align data sources when comparing them to the 1,590,982 non-retracted papers also cited on Wikipedia and indexed in Altmetric. The Crossref Event API proved impractical for extracting such a large volume of citations to non-retracted papers, so relying on Altmetric for both retracted and non-retracted papers ensured source consistency for this descriptive analysis. This dataset is used solely for comparing academic domain distributions and is not part of our modeling of correction outcomes. To enrich the dataset, we retrieved academic domain classifications from OpenAlex, along with additional metadata such as open access status, citation counts, and co-author team sizes. All data were collected in October 2024.

\subsection{Types of Retracted Paper Citations}
  
When retracted papers are first introduced as citations on Wikipedia, they can appear in the following ways: 

\begin{itemize}
\item{\textit{Cited Before Retraction.}}
These are instances where papers are cited on Wikipedia before their official retraction date. At the time of citation, the paper’s retracted status was not yet known, and no corrective context was provided in the article. If these citations remain after retraction without being updated or flagged, they risk misleading readers.

\item{\textit{Cited After Retraction Without In-Text Acknowledgment.}}
These citations occur after the paper has been officially retracted, but there is no indication of the retraction in the Wikipedia article. Readers may mistakenly assume that the paper's results are valid, which could lead to misconceptions.

\item{\textit{Cited After Retraction With In-Text Acknowledgment.}}
In some cases, retracted papers are introduced with explicit acknowledgment of their retraction status. This definition intentionally excludes automated warnings in the reference section, as our focus is on the human editorial actions that make a retraction's status transparent to a reader of the main article text. At the same time, we treat RetractionBot flagging as an observable feature that may be associated with subsequent reader-facing repair.
\end{itemize}

For this study, we consider problematic citations to be the ones \textit{Cited Before Retraction} and \textit{Cited After Retraction Without In-Text Acknowledgment}. These scenarios pose significant risks of misinformation, as readers may unknowingly rely on invalid research. We focus exclusively on whether retraction status is clearly mentioned in the article text. Merely critiquing a paper or casting doubt on its findings does not meet our threshold for acknowledgment unless the retraction is explicitly stated.

To identify these cases, we tracked when each retracted paper first appeared in its associated Wikipedia articles by searching for its DOI and title across revision histories. One of the authors then manually examined whether the retraction status was mentioned in the main text of those initial citing revisions.

\subsection{Handling Problematic Retracted Paper Citations}

By manually inspecting the latest version of pages containing problematic citations in our dataset, we distinguish three reader-relevant ways that retraction status may be handled in an article: 

\begin{itemize}
\item{\textit{Complete Removal.}}
The most straightforward approach to correcting citations of retracted papers has been to remove them entirely from the Wikipedia article. This method eliminates any potential for future misinformation stemming from the problematic source. However, it can be controversial if the retracted paper contributed historically significant or contextually relevant information to the article. In academic paper writing, the complete removal of retracted paper citations is often preferred when editors and reviewers do not consider its continued presence essential \cite{retractionwatch2018how}.

\item{\textit{In-Text Retraction Acknowledgment.}}
A second approach involves placing a clear warning near the citation in the Wikipedia article text, ideally in the same paragraph where the retracted paper is referenced. This method informs readers about the paper's retracted status while preserving the citation for transparency or historical context. Such warnings may appear as editorial notes or parenthetical remarks, e.g., ``(This paper has been retracted).'' This category also includes other explicit in-text editorial flags, such as the \texttt{{{better source needed}}} template. We treat such editorial flags as a form of reader-facing repair because they explicitly signal sourcing problems in the main text, even if they do not name the retraction. This strategy balances accuracy and informational completeness, but it relies heavily on editors' consistency and awareness.

\item{\textit{Reference Section Acknowledgment Only.}}
In many cases, a paper’s retracted status is only acknowledged in the reference section, without modifying the article’s main text. Because these reference-level warnings are less visible to casual readers—who may not consult the references—this practice leaves the main text without a visible retraction signal, which may increase the risk that readers interpret the cited claim as uncontested. As a result, potentially outdated sourcing can remain in the main text without a reader-facing warning in Wikipedia’s content. For example, automated tools like RetractionBot \cite{Headbomb2024Signpost} are designed to flag retracted papers on Wikipedia by adding annotations primarily to the reference section.
\end{itemize}

For our analysis, we consider citations to be \textit{corrected} if they fall under \textit{Complete Removal} or \textit{In-Text Retraction Acknowledgment}, as these approaches offer high visibility to readers.

To identify these two types of corrections, we employed a two-step process. First, to find instances of \textit{Complete Removal}, we searched the revision history of each article for the titles and DOIs of the retracted papers. We normalized DOI strings (e.g., URL prefixes and case) and used case-insensitive matching in wikitext to identify first appearance and subsequent absence. Second, for all remaining cases where the citation was still present, we manually reviewed the most recent revision to identify instances of \textit{In-Text Retraction Acknowledgment}, assessing whether the retraction status was explicitly mentioned in the main text.

\subsection{Analytical Approach: Survival Analysis}
Our analysis examines which observable features are associated with the timing of reader-facing corrections for problematic citations. A significant portion of these citations remained uncorrected at the end of our data collection period. This type of incomplete observation is known as right-censored data. Standard regression models are not suited to handle such data. Therefore, to properly account for these censored observations, we employ survival analysis, a method commonly used to model time-to-event data in online communities \cite{ma2017write, mindel2025timely}. This method enables us to model the time until a correction occurs (the ``event'') while appropriately incorporating information from both corrected and uncorrected (censored) citations.

\subsection{Dependent Variable: Time to Correction} 
Our survival model analyzes the time it takes for a problematic citation to be corrected on Wikipedia. The primary dependent variable is the duration, measured in days, from when a citation becomes ``at risk'' of correction-that is, after it is both present on Wikipedia and officially retracted—until the correction event occurs.

Crucially, the start of the risk period for each citation is defined as the later of two dates: the paper's official retraction date or the date the citation was first added to the Wikipedia page. This ensures that a citation is only considered ``at risk'' after it is both present on Wikipedia and officially retracted. As a data cleaning step, we excluded a small number of ``proactive corrections''—cases where a correction was made before the official retraction date. To avoid undefined values when durations are zero (e.g., when the citation and retraction occur on the same date) and for models that include $\log(t)$-based terms, all durations were adjusted by adding a small $\epsilon=10^{-3}$ days.

A correction event is defined as either (i) removal of the citation that is sustained under our three-revision tolerance rule, or (ii) the addition of an in-text retraction warning in the immediate context of the citation. For citations that remained uncorrected as of our data collection endpoint (October 24, 2024), they are treated as right-censored observations. Their duration is the time from the start of their risk period until this endpoint.

\subsubsection{Correction via In-Text Retraction Acknowledgment}
For citations that were eventually corrected by mentioning the retraction status in the main text of the article, we begin by identifying a retraction-related keyword in the final, corrected revision of the Wikipedia article. These may include general terms like “retracted” or “withdrawn,” or more specific phrases such as “The article was retracted on 13 August 2010.” Using this final revision, we select a representative keyword or phrase associated with the correction and trace backward through the revision history to identify the earliest revision where that keyword appeared. We select this phrase on a per-citation basis to match the local wording used for the correction, rather than relying on a single global keyword list.

To avoid mistaking short-term rewording or editing noise as correction reversals, we required that the identified correction keyword or phrase, once added, remain in subsequent revisions, allowing for a temporary disappearance of up to three consecutive revisions. We acknowledge that acknowledgments may be reworded or restructured across revisions; we discuss this measurement limitation and our validation-based corrections in Section~\ref{sec:validation} and Section~\ref{sec:assumptions}. This heuristic balances robustness against transient noise with sensitivity to meaningful editorial behavior. Because articles may cite multiple retracted papers, we manually verify that the keyword refers to the specific citation of interest. If needed, we extract a more precise phrase from the later revision that unambiguously refers to the retracted paper in question.

\subsubsection{Correction via Removal}
For citations corrected through removal, we apply a similar logic. A citation is considered removed when it is absent from both the article’s main text and its reference section. We identify the earliest revision in which the citation is no longer present and remains removed, again defined as allowing for up to three subsequent revisions where the citation might briefly reappear. 

\subsubsection{Assumptions}\label{sec:assumptions}
Our approach to identifying correction events, which is based on a dual-track search for either Complete Removal or In-Text Retraction Acknowledgment, relies on several working assumptions:
\begin{itemize}
    \item {Wikipedia editors typically choose either in-text retraction acknowledgment or removal as a correction strategy for retracted citations.}
    \item {Once corrected, a citation typically remains corrected in that same form throughout subsequent revisions, allowing for brief reversals consistent with our three-revision tolerance.}
    \item {The retraction-related keyword or phrase, once added, is typically consistent in subsequent revisions, though it may be reworded or replaced (e.g., synonym substitution or text restructuring). In our manual validation (Section~\ref{sec:validation}), we found that remaining mismatches primarily stem from such rewriting dynamics or shifts between In-Text Acknowledgment and Complete Removal; we corrected the identified cases in the final analytical dataset.}
    \item {The presence or absence of a citation is reliably reflected in the article’s main text and references.}
\end{itemize}
By applying this methodology, we estimate the time from retraction to correction for each citation, enabling us to assess both the speed and consistency of Wikipedia’s response to retracted science.

\subsubsection{Validation of correction timestamps.}
Because correction timestamps are automatically derived from revision histories using the heuristics above, we manually validated a subset of cases and corrected identified mismatches caused by editorial rewriting or strategy switches (e.g., acknowledgment vs.\ removal). We report the validation design, sample sizes, and accuracy rates in Section~\ref{sec:validation}. This validation targets the automatically inferred correction timestamps (from revision heuristics), rather than the separate manual classification of citation types.

\subsection{Independent Variables}

\begin{table*}[t]
\centering
\begin{tabular}{lccccc}
\toprule
 & \textbf{Mean ($\mu$)} & \textbf{Std Dev ($\sigma$)} & \textbf{Min} & \textbf{Max} & \textbf{Median} \\
\midrule
\multicolumn{6}{l}{\textit{Paper Features}} \\
\quad Open Access & 0.62 & 0.49 & 0 & 1 & 1 \\
\quad Academic Citation Count (at risk start) & 148.90 & 419.99 & 0 & 5444 & 46.00 \\
\quad Team Size & 6.35 & 4.65 & 0 & 28 & 5.00 \\
\quad Physical Sciences (domain) & 0.13 & 0.34 & 0 & 1 & 0 \\
\quad Life Sciences (domain) & 0.36 & 0.48 & 0 & 1 & 0 \\
\quad Social Sciences (domain) & 0.08 & 0.27 & 0 & 1 & 0 \\
\quad Tweet Mention Count (at risk start) & 346.15 & 2162.73 & 0 & 33946 & 1.50 \\
\addlinespace
\multicolumn{6}{l}{\textit{Retraction Event Features}} \\
\quad Time to Retraction (days) & 2414.00 & 2097.01 & 0 & 11344 & 1926.50 \\
\quad Pre-Retraction Editorial Event & 0.12 & 0.33 & 0 & 1 & 0 \\
\quad Retraction Reason (salient) & 0.51 & 0.50 & 0 & 1 & 1 \\
\addlinespace
\multicolumn{6}{l}{\textit{Wikipedia Page Features}} \\
\quad RetractionBot (ever during risk period) & 0.59 & 0.49 & 0 & 1 & 1 \\
\quad Number of Unique Editors (at risk start) & 152.13 & 275.55 & 0 & 2343 & 34.50 \\
\quad Controversy Noted at Initial Citation & 0.05 & 0.22 & 0 & 1 & 0 \\
\quad Media Count (at risk start) & 3.16 & 11.83 & 0 & 266 & 0 \\
\quad Categories Count (at risk start) & 3.43 & 3.92 & 0 & 56 & 2.00 \\
\quad Reference Count (at risk start) & 82.66 & 137.25 & 0 & 1416 & 36.00 \\
\bottomrule
\end{tabular}
\caption{Summary statistics for the independent variables. RetractionBot is modeled as a time-dependent covariate in the extended Cox model; here we report whether a citation was ever monitored during its risk period (0/1).}
\label{tab:summary_statistics}
\end{table*}

To identify factors that are associated with faster or slower correction of retracted paper citations on Wikipedia, we examine three different groups of characteristics, as follows (see Table \ref{tab:summary_statistics}):

\subsubsection{Paper Features}

Paper characteristics are particularly important because they capture intrinsic properties of the publications that may be associated with differences in how quickly citations are corrected on Wikipedia. These data were retrieved from external scholarly databases, including OpenAlex \cite{priem2022openalex}, Retraction Watch \cite{oransky2012retraction}, and Altmetric \cite{altmetric}.

\begin{itemize}
    \item \textit{Open Access.} Open access status is coded as a binary variable. This variable allows us to examine whether open-access availability is associated with correction timing on Wikipedia. Prior research suggests that open-access papers may receive greater visibility and more citations due to their unrestricted availability \cite{piwowar2018state,teplitskiy2017amplifying}. Zheng et al. (2024) found that open access may help speed up the detection and retraction of flawed articles in biochemistry \cite{zheng2024gold}.

    \item \textit{Academic Citation Count.} Academic citations serve as a proxy for a paper's scholarly impact and visibility. To avoid look-ahead bias, we measured this variable dynamically for each citation pair. Specifically, we only counted citations the paper had accumulated before the start of its risk period on Wikipedia, ensuring our measure of scholarly impact could not be influenced by the correction process itself. Using the OpenAlex API, we determined the number of citations a retracted paper had received up to the start of each citation's unique risk period (i.e., the point after it was both retracted and cited on Wikipedia). While a paper's higher citation count could increase its visibility and be associated with a faster correction on Wikipedia \cite{van2011science, wang2021science}, it could also create editorial inertia if its findings are widely accepted, resulting in a slower correction. 

    \item \textit{Team Size.} Team size is measured by counting the number of paper authors. Larger teams may contribute to greater research visibility and faster dissemination due to broader professional networks \cite{li2022untangling}. However, prior studies suggest that increased team size can also complicate accountability, potentially leading to delays in retraction or correction processes \cite{mongeon2016costly}.

    \item \textit{Academic Domains.} 
    We include academic domains to examine whether the speed of citation correction varies across broad fields of research. Domains are derived from OpenAlex’s \texttt{primary\_topic} field, which is based on Scopus’s ASJC classification \cite{Elsevier} and assigned using OpenAlex’s topic classification methods \cite{openalexend}. We focus on four high-level domains: Health Sciences, Life Sciences, Physical Sciences, and Social Sciences. To incorporate these categorical variables into our regression model, we applied one-hot encoding and omitted Health Sciences as the reference category to avoid multicollinearity. Differences across domains may reflect disparities in editorial practices, public scrutiny, or disciplinary norms. For example, Health Sciences often receive greater media and public attention, which can accelerate the detection and correction of retracted work \cite{furman2012governing}. In contrast, corrections in the Social Sciences may lag due to longer publication cycles.

    \item \textit{Tweet Mention Count.}
     Tweet mention count data are collected from Altmetric LLC \cite{altmetric}. To ensure methodological consistency and avoid look-ahead bias, we measured the number of tweets up to the start of each citation's unique “at risk” period. This count serves as a proxy for the paper’s online visibility and public exposure at the moment it became problematic on Wikipedia. While a higher count of these pre-existing tweets could signal a more prominent paper that editors are more likely to scrutinize, we acknowledge that the attention generated by the retraction event itself is another important mechanism. Although modeling this second effect is a valuable direction for future work, this study focuses on the role of a paper's pre-existing visibility. Prior research highlights the role of online engagement in scientific controversies, finding that public attention on platforms like Twitter often spikes around the time a paper is retracted \cite{peng2022dynamics}. While news and blog mentions also serve as valid proxies for online engagement, we found that they were highly correlated with tweet mention counts. To avoid multicollinearity in our model, we therefore selected the Tweet mention count as the most representative measure.

\end{itemize}

\subsubsection{Retraction Event Features}

This feature group captures characteristics of the retraction event itself. These features provide context on the severity of the paper's issues and the timeline of its withdrawal, which may be associated with how quickly citations are corrected. Data for these features were retrieved from OpenAlex \cite{priem2022openalex} and Retraction Watch \cite{oransky2012retraction}.

    \begin{itemize}
    \item \textit{Time to Retraction.} Time to retraction is calculated as the time between a paper's publication date and its official retraction date, retrieved from OpenAlex and Retraction Watch. Longer times between publication and retraction may be associated with slower correction, as outdated citations may become embedded in the platform's content. Prior studies have highlighted the negative impact of delayed retractions on scientific credibility and public trust \cite{brainard2018massive}.

    \item \textit{Pre-Retraction Editorial Events.} Editorial events include corrections, expressions of concern, or other non-retraction updates associated with a given DOI. If such events existed, the variable was labeled as 1. Such events could serve as warnings for editors or readers, potentially influencing how quickly problematic citations are identified and addressed.

    \item \textit{Retraction Reason.} According to the framework presented in \citet{zhang2020collaboration}, the retraction reason can be categorized into 8 groups: ``plagiarism'', ``self-plagiarism'', ``falsification/manipulation'', ``error \& unreliable results'', ``authorship issues'', ``ethical issues'', ``others'', and ``not available/lack of information''. For this study, we focused specifically on falsification/manipulation (fraud) and error \& unreliable results (error). We prioritized these categories because they most directly undermine the epistemic validity of the scientific claims. In contrast, retractions due to authorship problems, ethical issues, or other administrative errors, while serious violations of academic norms, do not necessarily imply that the scientific findings themselves are factually incorrect. Following this framework, we searched for the keywords ``falsification,'' ``manipulation,'' and ``misconduct'' in the Retraction Watch reasons to identify fraud papers, and we used the keywords ``error,'' ``contamination,'' and ``unreliable'' to identify error papers. These fraud and error categories were then combined into a broader category labeled ``fraud \& errors,'' capturing retractions that may draw more editorial attention. Papers falling into this category were assigned a binary value of 1.
    \end{itemize}

\subsubsection{Wikipedia Page Features} 
Following prior work on language-agnostic Wikipedia article features \cite{das2024language}, we include a set of page-level characteristics measured at the start of each citation’s risk period. This feature group captures characteristics of Wikipedia pages associated with variation in the timing of reader-facing corrections. We aim to understand how Wikipedia's collaborative and dynamic editorial processes are associated with the persistence of problematic citations. All features in this group are derived from Wikimedia data, retrieved through the Wikimedia REST \cite{MediaWiki} and MediaWiki Action \cite{WikiAction} APIs. 
\begin{itemize}
    \item \textit{RetractionBot Intervention.} We treat the presence of a RetractionBot warning as a time-dependent covariate. We recorded the exact timestamp when the bot first flagged the citation pair. In our time-varying covariate construction, the status is 0 before this timestamp and 1 after for the citation pair. This specification allows us to estimate how the hazard of correction differs before versus after a RetractionBot flag.
    
    \item \textit{Number of Unique Editors.} This variable serves as a proxy for the level of collaborative oversight on a given Wikipedia page. To ensure this was a valid baseline covariate and to avoid look-ahead bias, we measured it dynamically. For each citation pair, we calculated the number of distinct human editors who had contributed to the page up to the start of that citation's risk period (i.e., the point after the cited paper was both retracted and present on the page). To ensure the measure reflects meaningful editorial engagement, we excluded bot and anonymous edits from this count. The rationale is that a page with a larger community of editors at the time a citation becomes at risk may experience greater collaborative oversight, potentially correlating with faster correction.

    \item \textit{Controversy Noted at Initial Citation.} This feature indicates whether the Wikipedia article describes the cited paper as controversial or disputed at the time it was first added to the page. We identify this by manually checking whether the initial citation frames the paper as debated, flawed, or otherwise questionable. Such editorial framing may capture early skepticism and be associated with earlier correction.
    
    \item \textit{Reference Count.} We use the number of references on the Wikipedia page as a proxy for sourcing density and page complexity. To avoid look-ahead bias, we measure this feature dynamically at the start of each citation’s risk period. Pages with more references may reflect stronger sourcing norms and editorial attention, but they may also impose a higher maintenance burden, potentially slowing the identification and correction of problematic citations.

    \item \textit{Categories Count.} This feature captures how broadly the Wikipedia page is situated within the platform’s topical classification system, measured as the number of categories assigned to the page at the start of the citation’s risk period. A larger number of categories may increase the likelihood that the page is monitored by multiple interest groups or WikiProjects, potentially correlating with earlier detection and correction.

    \item \textit{Media Count.} We measure the number of media items (e.g., images and other embedded files) included on the page at the start of the citation’s risk period. Media-rich pages may attract more readership and editorial attention, which may be associated with earlier correction, although they may also be longer or more complex.

\end{itemize}
We excluded revision count at risk time from the model because it was nearly perfectly correlated with the number of unique editors at risk time. We omit other size proxies (e.g., page length, headings, and wikilinks) because they are highly collinear with these structural measures and would add redundancy.

\subsection{Data Preprocessing and Model Specification}

Before model fitting, we preprocessed the independent variables to ensure model stability. We applied a log-transformation $\log(x+1)$ to right-skewed count-like variables, including academic citation count, team size, tweet mention count, time to retraction, and number of unique editors. Subsequently, these continuous variables were standardized using z-score normalization to facilitate comparison of coefficient magnitudes (i.e., log-hazard ratios) across covariates. Binary and dummy variables were left unchanged. Our initial sample of 845 problematic citations was reduced to a final analytical sample of 762 after excluding 83 ``proactive corrections''—cases where the correction was made before the official retraction date. We exclude these cases because our risk-set definition begins only once a paper is officially retracted; corrections made prior to retraction likely reflect a different process (e.g., pre-retraction controversy coverage) and would yield negative durations under this operationalization.

Our modeling strategy proceeded in three steps: (1) constructing a time-dependent covariate for RetractionBot flagging (0 before the first flag, 1 after) to avoid immortal time bias, (2) assessing the proportional hazards assumption using Schoenfeld residual tests (global and covariate-specific) \cite{schoenfeld1982partial}, and (3) specifying an extended Cox model with covariate–$\log(t)$ interaction terms for variables that violated proportional hazards.

\subsubsection{Construction of Time-Dependent Covariate: RetractionBot}
A critical methodological challenge in analyzing the association of RetractionBot flagging with correction timing is the temporal nature of its intervention. Treating the bot as a static binary variable would introduce immortal time bias \cite{suissa2008immortal}, as citations must ``survive'' (remain uncorrected) long enough to be visited by the bot. To rigorously address this, we structured our dataset using the counting process format (start-stop format) \cite{therneau2000modeling}. We split the observation period of any citation targeted by the bot into two distinct intervals:
\begin{itemize}
  \item \textit{Pre-intervention interval} $(0, t_{\mathrm{bot}}]$ where the citation is treated as unexposed ($RetractionBot=0$).
  \item \textit{Post-intervention interval} $(t_{\mathrm{bot}}, t_{\mathrm{end}}]$ where it is treated as exposed ($RetractionBot=1$).
\end{itemize}

Here, $t_{\mathrm{bot}}$ is the timestamp of the first RetractionBot flag for the citation pair, and $t_{\mathrm{end}}$ is the event time (correction) or the censoring time (October 24, 2024), whichever comes first.
This transformation allows the risk set to be updated instantaneously at the moment of the bot's intervention. Citations never visited by the bot contribute a single interval with $RetractionBot=0$ throughout follow-up. After splitting, the analytical dataset contains 1,154 start–stop intervals (rows), representing 762 unique citation pairs, with 484 observed correction events.

\subsubsection{Assessment of Proportional Hazards and Model Refinement}

We adopted a rigorous, iterative approach to assess the proportional hazards (PH) assumption, ensuring our final model specification accurately reflected the underlying data structure. We assessed the proportional hazards assumption using scaled Schoenfeld residual tests implemented in R (\texttt{survival::cox.zph}), reporting both global and covariate-specific tests.

\begin{itemize}

\item \textit{Step 1: Initial Diagnostics.} We first fitted a Cox model incorporating the time-dependent RetractionBot intervention and all static variables, using one-hot encoding for Academic Domains (using Health Sciences as reference). Diagnostic tests based on scaled Schoenfeld residuals (see Appendix Table \ref{tab:ph_test_first}) revealed severe violations of the PH assumption. Specifically, the \textit{Academic Domain} variables ($p < 0.001$), \textit{Tweet Mention Count} ($p < 0.001$), \textit{Open Access} ($p < 0.001$), and \textit{Retraction Reason (Salient) ($p < 0.001$)} showed strong deviations. Furthermore, we observed likely spurious violations in several other variables (e.g., \textit{Wikipedia Page Reference Count}, \textit {Categories Count}, and \textit{Media Count}), suggesting that unmodeled heterogeneity in baseline hazards across disciplines may be distorting the model estimates.

\item \textit{Step 2: Stratification.} To address the violations related to disciplinary differences, we stratified the model by \textit{Academic Domain}. This modification allows each discipline to have a distinct baseline hazard function while sharing common coefficients for other covariates. Re-evaluation of the PH assumption on this stratified model (see Appendix Table \ref{tab:ph_test_second}) confirmed the effectiveness of this approach: violations for control variables like \textit{Media Count} ($p = 0.569$) and \textit{Categories Count} ($p = 0.1345$) were resolved.

\item \textit{Step 3: Time-Dependent Coefficients.} However, the stratified diagnostics confirmed that \textit{Tweet Mention Count} ($p < 0.001$), \textit{RetractionReason} ($p < 0.05$), and \textit{Open Access} ($p < 0.05$) continued to significantly violate the PH assumption. This indicates that the effects of social media attention and page sourcing density are not constant but evolve over the risk period. To model these dynamic effects, we specified the final Extended Cox model by including interaction terms for both variables with the logarithm of time ($TweetCount \times \ln t$ and $ReferenceCount \times \ln t$).

\end{itemize}

\subsubsection{Extended Cox Model Specification}

The hazard rate for citation $i$ at time $t$ is modeled as:
\begin{align}
h_i(t) = h_{0g}(t)\exp\Big(
&\beta_1 \,\mathrm{RetractionBot}_i(t) \nonumber \\
&+ \beta_2\, \mathrm{TweetCount}_i + \gamma_2\, \mathrm{TweetCount}_i \ln(t+\epsilon) \nonumber \\
&+ \beta_3\, \mathrm{OpenAccess}_i + \gamma_3\, \mathrm{OpenAccess}_i \ln(t+\epsilon) \nonumber \\
&+ \beta_4\, \mathrm{TeamSize}_i
+ \beta_5\, \mathrm{TimeToRetraction}_i
+ \beta_6\, \mathrm{PreRetractionEditorial}_i \nonumber \\
&+ \beta_7\, \mathrm{RetractionReason}_i + \gamma_7\, \mathrm{RetractionReason}_i \ln(t+\epsilon) \nonumber \\
&+ \beta_8\, \mathrm{CitationsAtRisk}_i
+ \beta_9\, \mathrm{ReferenceCount}_i
+ \beta_{10}\, \mathrm{UniqueEditors}_i \nonumber \\
&+ \beta_{11}\, \mathrm{ControversyNoted}_i
+ \beta_{12}\, \mathrm{MediaCount}_i
+ \beta_{13}\, \mathrm{CategoriesCount}_i
\Big).
\end{align}

\noindent where:

\begin{itemize}
    \item $h_{0g}(t)$ is the stratified baseline hazard for academic domain $g$ (Health, Life, Physical, or Social Sciences), allowing baseline risk shapes to vary across domains.
    \item $t$ denotes time since the start of the risk period (in days). To avoid $\ln(0)$, we use $\ln(t+\epsilon)$ with $\epsilon=10^{-3}$.
    \item $\mathrm{RetractionBot}_i(t)$ is a time-dependent covariate indicating whether RetractionBot has become active for citation $i$ by time $t$ (0 before activation, 1 after).
    \item $\mathrm{TweetCount}_i$ is the tweet mention count at the start of the risk period. Its effect is allowed to vary over time via the interaction $\mathrm{TweetCount}_i \times \ln(t+\epsilon)$.
    \item $\mathrm{OpenAccess}_i$ indicates whether the cited paper is open access. We allow this association to vary over time via $\mathrm{OpenAccess}_i \times \ln(t+\epsilon)$.
    \item $\mathrm{TeamSize}_i$ is the number of authors of the cited paper.
    \item $\mathrm{TimeToRetraction}_i$ is the number of days from publication to retraction.
    \item $\mathrm{PreRetractionEditorial}_i$ indicates whether the paper received an editorial notice prior to retraction (e.g., an expression of concern).
    \item $\mathrm{RetractionReason}_i$ is an indicator for \textit{fraud/misconduct/error/unreliable results} (1) vs.\ \textit{other reasons} (0). Its effect is allowed to vary over time via $\mathrm{RetractionReason}_i \times \ln(t+\epsilon)$.
    \item $\mathrm{CitationsAtRisk}_i$ is the academic citation count at the start of the risk period.
    \item $\mathrm{ReferenceCount}_i$ is the number of references on the Wikipedia page at the start of the risk period.
    \item $\mathrm{UniqueEditors}_i$ is the number of unique editors contributing to the Wikipedia page (measured at risk start).
    \item $\mathrm{ControversyNoted}_i$ indicates whether the citation was initially framed as controversial.
    \item $\mathrm{MediaCount}_i$ and $\mathrm{CategoriesCount}_i$ are Wikipedia page structural features (measured at risk start).
\end{itemize}

We estimated models using robust (sandwich) standard errors \cite{lin1989robust} clustered by citation pair to account for within-pair dependence introduced by the start–stop (counting-process) representation. 

\subsubsection{Implementation}
We estimated the extended Cox models in R using \texttt{survival::coxph()} with a counting-process (start–stop) specification and robust standard errors clustered at the citation-pair level. Kaplan–Meier curves and survival descriptives were generated and visualized in Python using the \texttt{lifelines} package.

\section{Results}
\subsection{Overview of Retracted Papers on Wikipedia}

\begin{figure}[t]
\centering
\includegraphics[width=0.7\columnwidth]{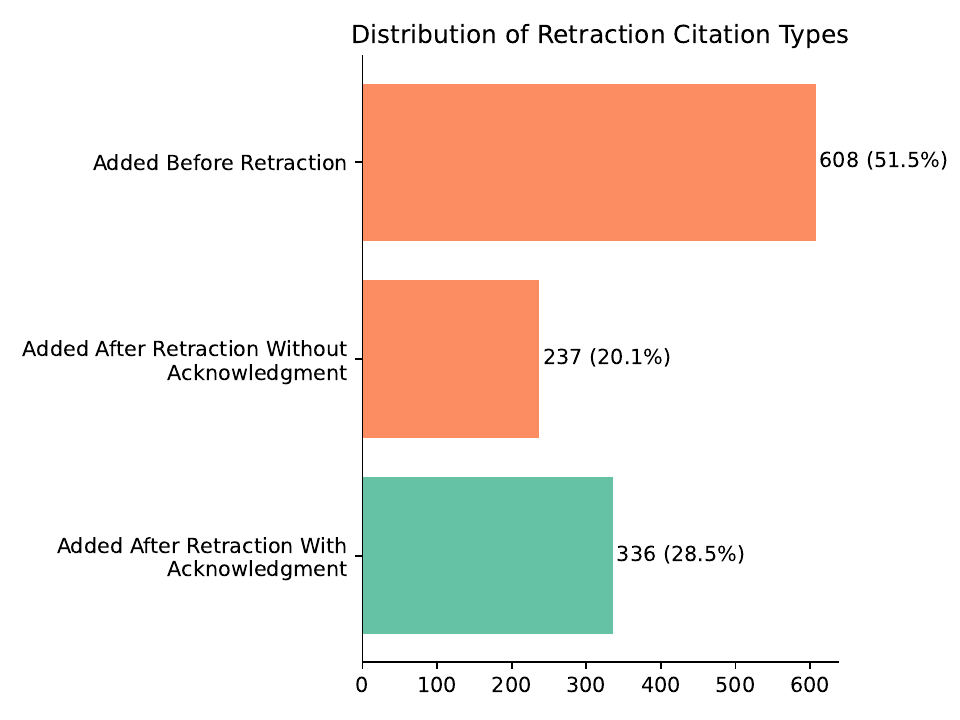} 
\caption{Share of problematic (orange) vs unproblematic (green) retracted-paper citations on Wikipedia.
}
\label{distribution}
\end{figure}

\subsubsection{Problematic Citations}
 We analyze the distribution of different types of retracted paper citations on Wikipedia. As shown in Figure \ref{distribution}, among the 1,181 citation pairs found, 845 were identified as problematic citations, with 608 cited before retraction and 237 cited after retraction without proper in-text warnings. This means that 71.6\% of all collected citations represent problematic cases, where retracted papers were either cited before their retraction status became known or cited afterwards without adequate warnings to readers. 51.5\% of all citations fall into the former category. These citations often persist in the absence of systematic monitoring, leaving articles exposed to outdated or invalidated information. In contrast, 28.4\% of the citations were non-problematic, meaning that their retracted status was appropriately mentioned at the time of citation. Critically, the category of citations introduced after a paper's retraction but without proper in-text warnings constitutes 20.1\% of all citations in our sample, revealing the limitations of current editorial and automated interventions in flagging and fixing such citations. 

 \begin{figure}[t]
\centering
\includegraphics[width=0.75\columnwidth]{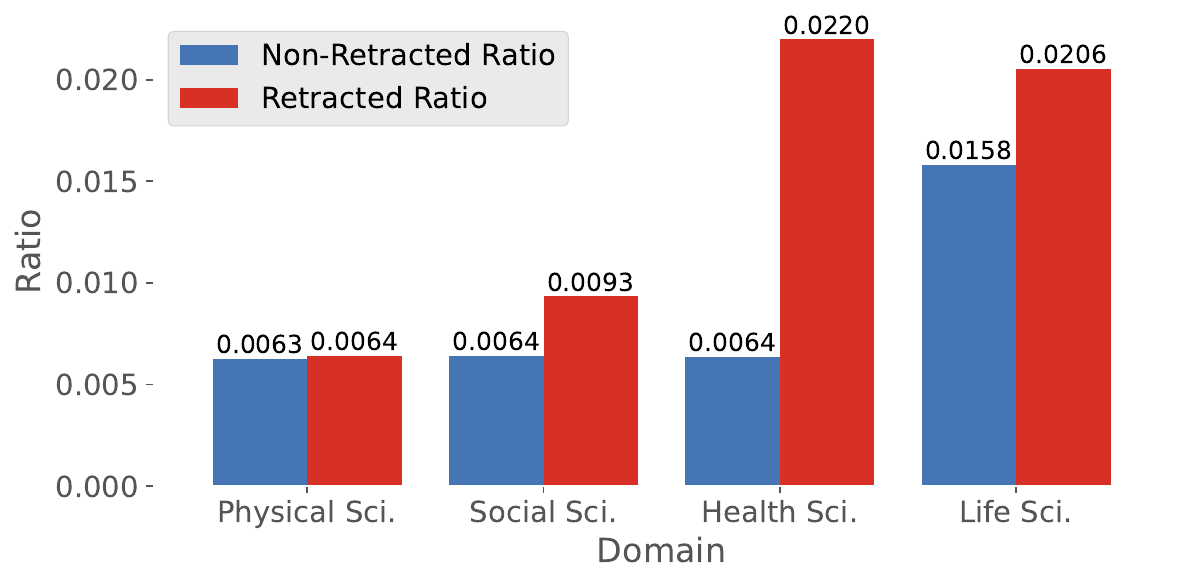} 
\caption{Comparison of the ratios of retracted and non-retracted papers cited on Wikipedia across academic domains. The retracted ratio represents the fraction of all retracted papers within a domain that are cited on Wikipedia (blue). In contrast, the non-retracted ratio is the fraction of non-retracted papers cited on Wikipedia (red).}
\label{domain}
\end{figure}

\subsubsection{Citation of Retracted vs Non-retracted Papers Across Academic Domains}
We examined how retracted papers are cited on Wikipedia relative to non-retracted papers across major academic domains. We compute the retracted ratio as the proportion of retracted papers within each domain that Wikipedia cites. Similarly, the non-retracted ratio is calculated as the proportion of non-retracted papers within each domain that Wikipedia cites. For example, in Life Sciences, the retracted ratio refers to the fraction of all retracted Life Sciences papers that appear as citations on Wikipedia. This ratio is 0.0206, representing the fraction of all retracted Life Sciences papers appearing as Wikipedia citations. We used a separate dataset acquired from Altmetric for this analysis, and we collected 538 retracted papers indexed in Altmetric. Figure \ref{domain} compares ratios between retracted and non-retracted papers across four academic domains: Physical Sciences, Social Sciences, Health Sciences, and Life Sciences.

The Health Sciences domain stands out with the highest ratio of retracted papers cited on Wikipedia, at 0.0220, compared to a much lower 0.0064 ratio for non-retracted papers. This disparity suggests a heightened vulnerability of this domain to problematic citations. The retracted-to-non-retracted citation ratio is 3.44 times higher in Health Sciences (0.0220 vs. 0.0064), indicating an overrepresentation of retracted work among Wikipedia-cited papers in this domain.

In contrast, the Physical and Social Sciences exhibit near parity between the ratios of retracted and non-retracted papers cited (0.0064 vs. 0.0063 for Physical Sciences; 0.0093 vs. 0.0064 for Social Sciences), suggesting more balanced editorial attention. Life Sciences show a modest gap, with 0.0206 for retracted and 0.0158 for non-retracted papers, highlighting that while problematic citations are prevalent, they are not disproportionately overrepresented relative to overall citation activity in the domain.

These findings reveal that citation persistence varies by academic field, with Health Sciences exhibiting the most acute overrepresentation of retracted work. Given Wikipedia’s widespread use as a public-facing medical reference, addressing this gap is particularly crucial for maintaining information integrity.

\subsection{Correcting Problematic Citations on Wikipedia}

\begin{figure}[t]
\centering
\includegraphics[width=0.7\columnwidth]{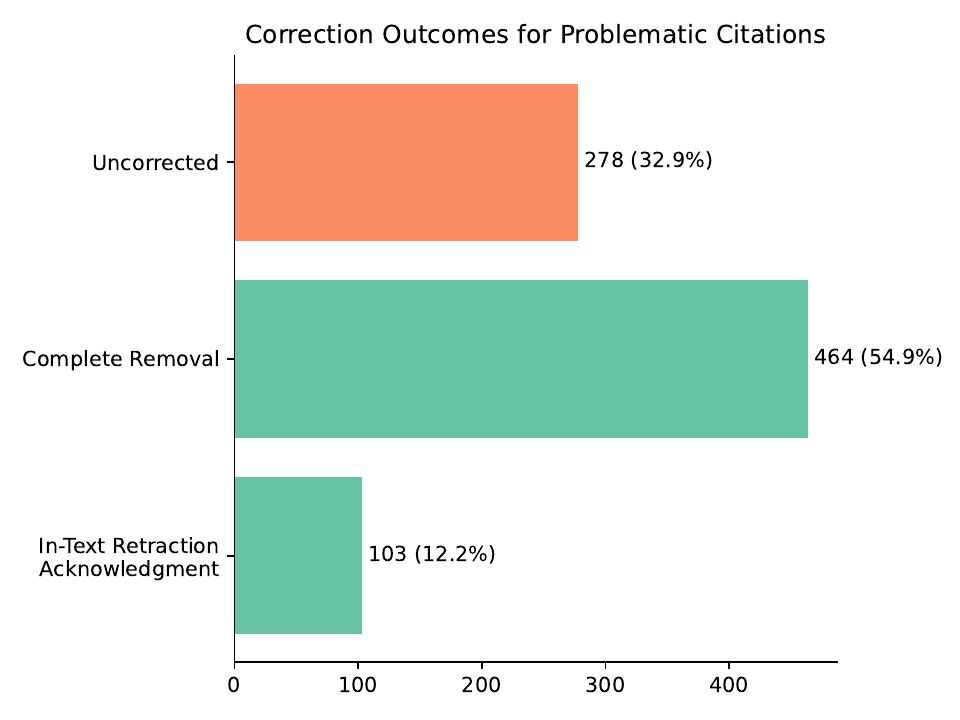} 
\caption{Distribution of corrections of problematic citations on Wikipedia.
}
\label{correction}
\end{figure}

\subsubsection{Distribution of the Corrections}
As shown in Figure \ref{correction}, 32.9\% of problematic retracted paper citations on Wikipedia ($n=278$) remained uncorrected when the data was collected, exposing readers to possible misinformation. While 12.2\% are addressed through in-text retraction acknowledgments ($n=103$) and 54.9\% are completely removed ($n=464$), the non-negligible percentage of unaddressed cases reveals substantial gaps in editorial intervention. 

\subsubsection{Validation of Time to Correction}\label{sec:validation}
To assess the accuracy of our automatically derived correction timestamps, we manually validated the identified correction revisions against human judgments. For in-text retraction acknowledgments ($n=103$), one author validated all cases; for complete removals ($n=464$), we validated a random sample of 100 cases. Concretely, the author opened the revision history for each sampled article–citation pair, located the first revision where the correction is visibly present (either the first stable in-text retraction acknowledgment or the first stable removal of the citation), and compared that revision ID/timestamp to the one returned by our automated pipeline. When the two did not match, the author reviewed adjacent revisions to determine whether the discrepancy was due to rewording of the acknowledgment token/phrase or a switch between correction strategies.

Under our validation criterion—whether the automatically identified correction revision matched the manually identified correction revision (allowing for our three-revision tolerance for brief reversals)—the agreement rate was 87\% for in-text acknowledgments and 94\% for complete removals. Disagreements primarily arose from editorial rewriting that altered the specific token or phrase used to signal retraction status, or from strategy switches between in-text acknowledgment and complete removal across revisions. We manually corrected these cases in the final analytical dataset used for subsequent analyses.

\begin{figure}[t]
\centering
\includegraphics[width=0.8\columnwidth]{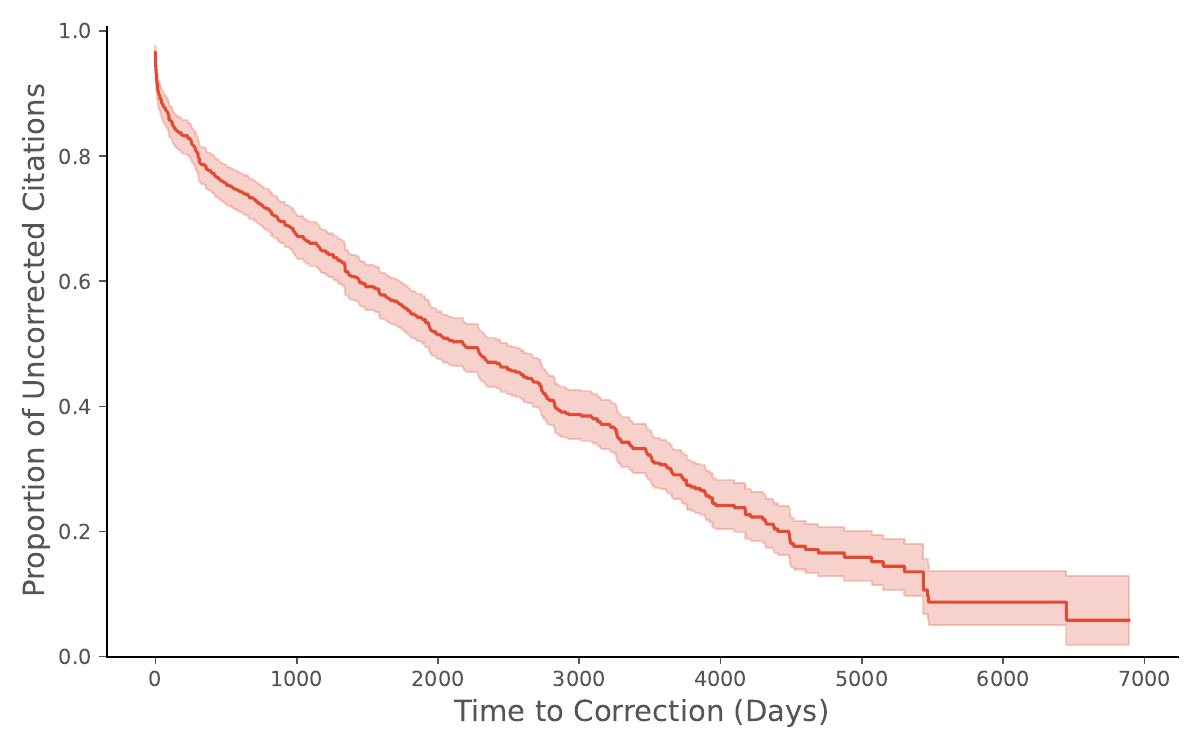} 
\caption{Kaplan-Meier survival curve of problematic citations. The y-axis shows the proportion of citations that remain uncorrected over time. The x-axis represents the time to correction in days, measured from the start of the risk period for each citation, which is defined as the later of the paper's official retraction date or the citation's addition to Wikipedia. Shaded regions represent 95\% confidence intervals.}
\label{delay}
\end{figure}

\begin{figure}[t]
\centering
\includegraphics[width=0.8\columnwidth]{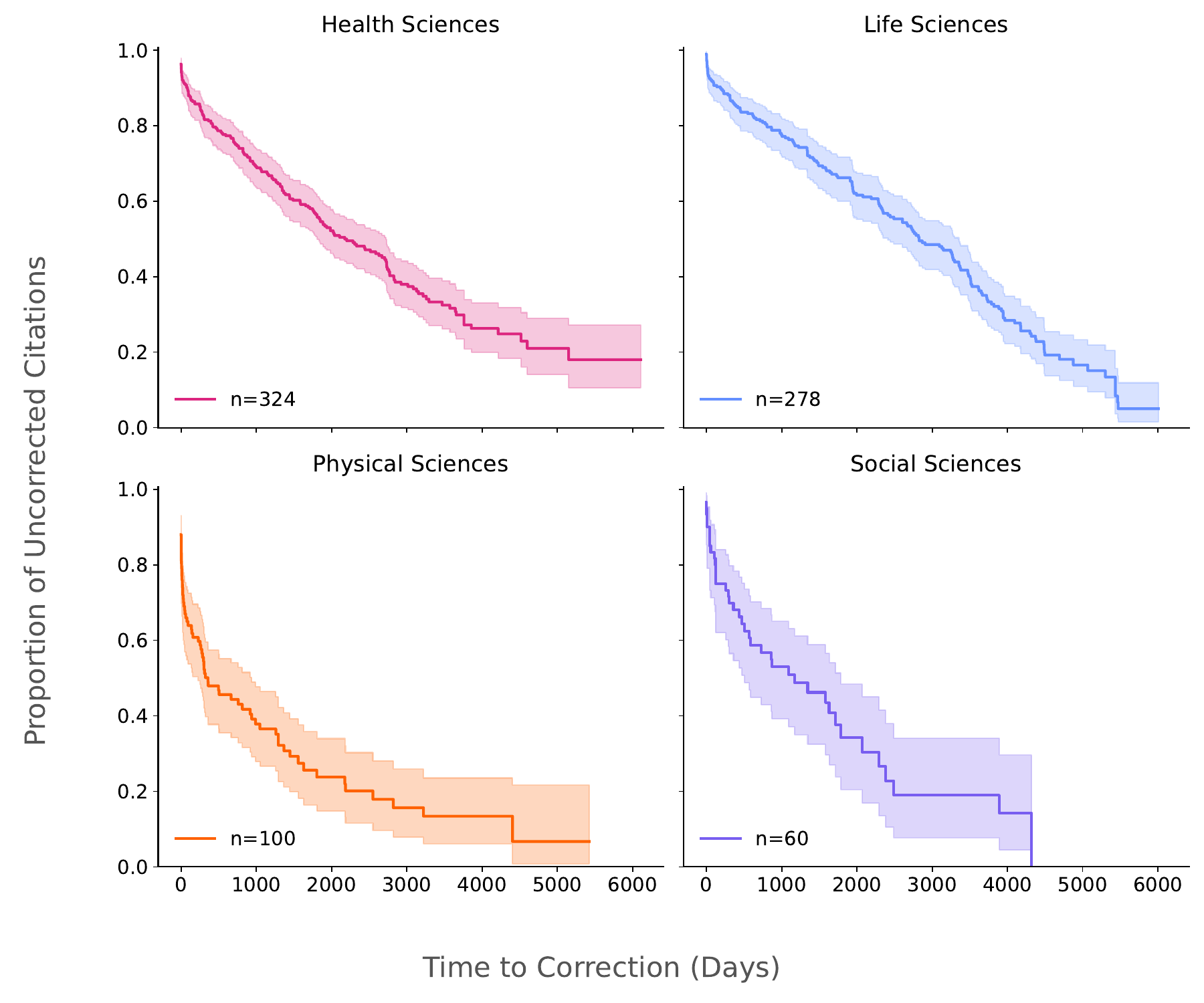} 
\caption{Kaplan--Meier survival curves showing the persistence of problematic retracted paper citations, stratified by academic domain. Curves are shown for descriptive comparison across fields. Domain sample sizes vary, particularly for Social Sciences. Shaded regions represent 95\% confidence intervals.}
\label{delay_domain}
\end{figure}

\subsubsection{Descriptive Analysis of Correction Times}
Our final analytical sample consists of 762 problematic citations. Of these, 484 were corrected during our observation period (the event), while 278 remained uncorrected and were right-censored at the end of the study. To visualize this process, we use the Kaplan-Meier estimator \cite{kaplan1958nonparametric}, a standard non-parametric method for plotting the probability of an event not yet occurring over time while properly accounting for censored data. The resulting survival curve for our sample is shown in Figure \ref{delay}.

The curve reveals a rapid rate of correction shortly after a citation's source paper has been retracted and is present on a Wikipedia page, followed by a long tail of persistence for those citations that are not addressed quickly. The median time to correction for problematic citations was found to be 1,344 days (approximately 3.68 years). While many corrections occur swiftly, the distribution is highly skewed. The longest observed duration for a citation to remain uncorrected before being censored was 6,107 days (over 16.73 years). Notably, the Kaplan-Meier analysis estimates that the probability of a citation remaining uncorrected after 1,000 days (approximately 2.74 years) is 67.3\%, indicating persistent challenges in timely intervention.

To examine whether this temporal pattern varies across knowledge domains, we next visualize Kaplan-Meier survival curves stratified by academic field. Figure \ref{delay_domain} shows clear heterogeneity in correction dynamics across domains.

Citations in the Physical Sciences ($n=100$) exhibit the fastest decline in survival probability, with a median time to correction of approximately 228 days, indicating more rapid correction. In contrast, the Health Sciences ($n=324$) and Life Sciences ($n=278$) display substantially longer persistence and a more pronounced long tail, with median correction times of 1,846 days and 2,720 days, respectively. Social Sciences ($n=60$) show intermediate behavior, with a median of 724 days; greater stepwise variation reflects the smaller sample size. These differences in correction dynamics across academic domains were statistically significant (Log-rank test, $\chi^2(3) = 77.14, p < 0.001$).

\begin{figure*}[t]
\centering
\includegraphics[width=1\textwidth]{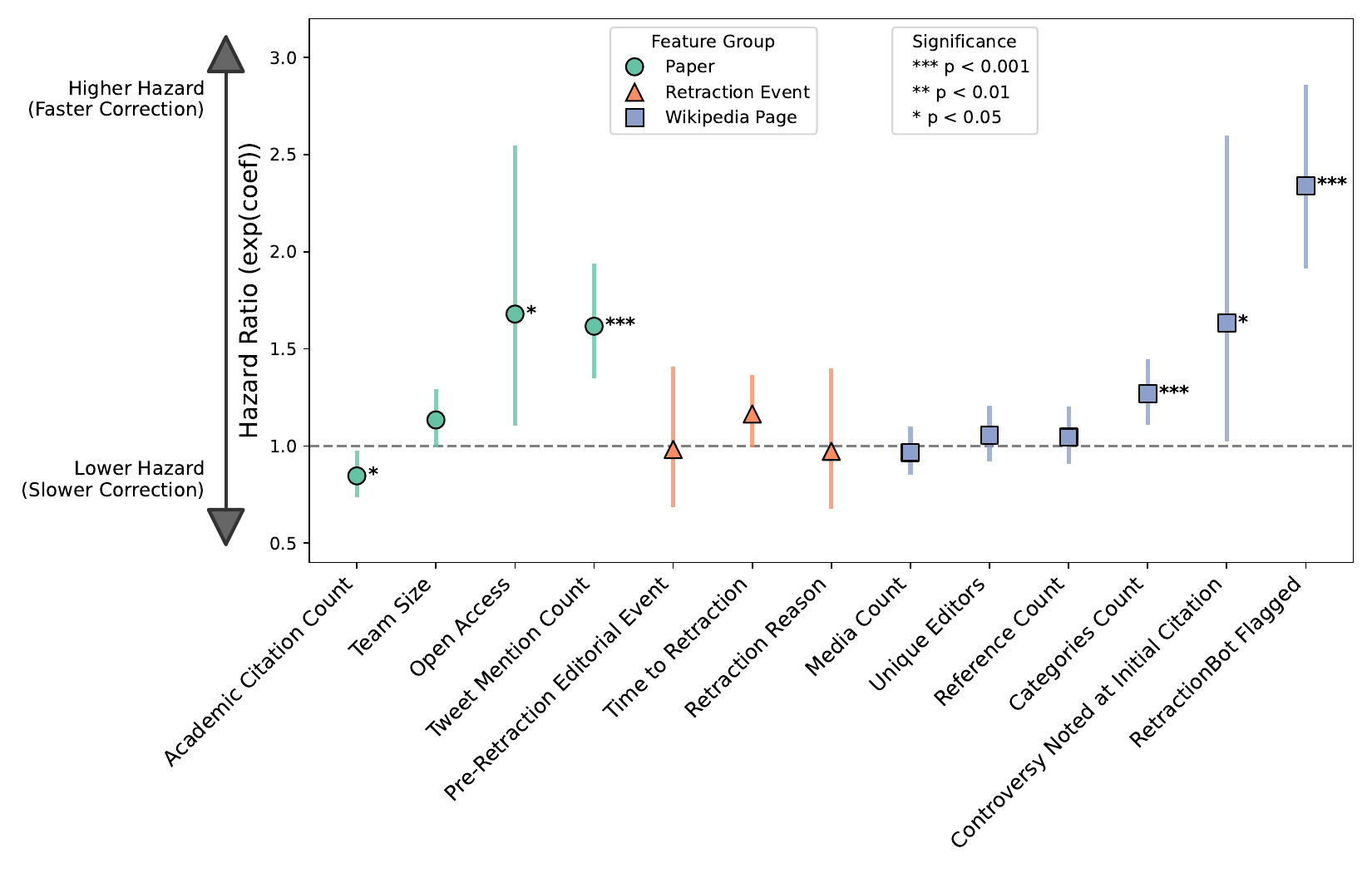} 
\caption{Hazard Ratios from the extended Cox Proportional Hazards model predicting the time to correction for problematic citations (Concordance = 0.684; log-likelihood ratio test: $p < 0.001$). 
Feature categories group the predictors. 
Points indicate the exponentiated coefficient estimates (Hazard Ratios); error bars represent 95\% confidence intervals. 
The shapes and colors distinguish feature groups. Note that for variables exhibiting non-proportional hazards, only the main effect coefficients are visualized here; significant time-dependent interaction terms are detailed in the main text. See Appendix Table \ref{tab:results} for full model results. Hazard Ratios $> 1.0$ are associated with a higher hazard of correction (i.e., corrected faster), while Hazard Ratios $< 1.0$ indicate a lower hazard (i.e., the problematic citation persists longer). For continuous predictors, hazard ratios correspond to a one–standard deviation increase in the log-transformed covariate (log1p, then z-scored).}
\label{regression}
\end{figure*}

\subsubsection{Associations with Time to Correction: Extended Cox Model Results}

To identify factors associated with the time until a problematic citation is corrected, we fitted an extended Cox proportional hazards model. This approach models the hazard (instantaneous rate) of correction, where a Hazard Ratio (HR) greater than 1 indicates a faster correction (higher likelihood of correction), whereas an HR less than 1 indicates a slower correction (persistence). For continuous predictors, hazard ratios correspond to a one–standard deviation increase in the log-transformed covariate (log1p, then z-scored). The overall model is statistically significant ($\text{log-likelihood ratio test}: p < 0.001$) and demonstrates good predictive accuracy ($concordance = 0.684$). We formally assessed multicollinearity using the diagnostic method proposed by Lee and Weissfeld (1996) for the extended Cox proportional hazards model with time-dependent covariates \cite{yul1996multicollinearity}. This diagnostic was computed from the model-based (non-robust) information matrix, as it is intended to evaluate linear dependence in the covariate design rather than sampling variability of the coefficient estimates. Robust (sandwich) standard errors are reported for inference, but they do not alter the collinearity structure of the covariates. The analysis yielded a maximum Condition Index (CI) of 2.87, well below the recommended threshold of 10. Variance decomposition proportions revealed no coupled loadings across predictors (see Supplementary Table \ref{tab:CI}), suggesting that the coefficient estimates are not materially affected by multicollinearity. Figure \ref{regression} summarizes the hazard ratio estimates, with full model results reported in Appendix Table \ref{tab:results}. 

\paragraph{Paper Features}
The visibility and accessibility of a paper are significantly associated with its correction hazard. Consistent with our expectations, papers that are Open Access are corrected at a significantly higher rate than those behind a paywall (HR $= 1.68$, $p < 0.05$). We find no evidence that this association changes over time (interaction $p = 0.778$). Similarly, a higher level of Twitter/X attention prior to retraction is linked to a significantly faster correction time (HR $= 1.62$, $p < 0.001$), suggesting that public attention on social media serves as an effective signal to Wikipedia editors. Notably, we found this association to be time-dependent (Interaction HR $= 0.97$, $p < 0.05$): the positive association between tweet volume and correction hazard attenuates as the risk period lengthens.

Conversely, a higher academic citation count at the risk start is significantly associated with a slower correction (HR $= 0.85$, $p < 0.05$). This aligns with the idea that highly cited papers may possess an ``authority shield'' that creates editorial inertia, making them harder to correct even after retraction. 

\paragraph{Retraction Event Features}
Time to retraction (publication-to-retraction duration) does not show a statistically significant association with correction timing ($p = 0.063$). In addition, the presence of a pre-retraction editorial event (e.g., an expression of concern) is not significantly associated with correction timing ($p = 0.921$). Retraction reason does not show a constant (time-invariant) association ($p = 0.881$); however, its association varies over time (interaction HR $= 1.07$, $p < 0.05$), indicating that differences by reason become more pronounced later in the risk period. Overall, these results suggest that retraction-event characteristics show limited average associations with reader-facing correction timing, with some effects emerging in a time-dependent manner.

\paragraph{Wikipedia Page Features}
Not all retractions are treated equally on Wikipedia. Citations flagged by RetractionBot exhibited the highest hazard of correction (HR $= 2.34$, $p < 0.001$), highlighting the role of bot-assisted maintenance in accelerating reader-facing repair. 

We find that the structural organization of a Wikipedia page is a more reliable predictor of the correction hazard than the sheer volume of editorial activity. We used categories count as a proxy for a page's integration into Wikipedia's knowledge graph and found that retracted citations on pages with more categories are associated with a significantly higher hazard of correction (HR $= 1.27$, $p < 0.001$).

Explicit signals of dispute were also strongly associated with the hazard of correction. Citations that were flagged as controversial at the time of addition had a significantly higher hazard of correction (HR $= 1.63$, $p < 0.05$). This suggests that visual cues of disagreement successfully attract attention to specific bibliographic errors. However, none of the number of unique editors ($p = 0.436$), the number of media files ($p = 0.614$), or the number of references ($p = 0.534$) was significant. This indicates that simply having a larger crowd of general contributors does not guarantee the specific form of scrutiny required to identify and fix retracted citations.

\section{Discussion}

The persistence of retracted paper citations on Wikipedia highlights a coordination vulnerability in how epistemic risks are handled within a large-scale coordination platform. Rather than directly observing editorial deliberation or decision-making, our analysis examines the outcomes of collaborative maintenance as recorded through revision histories. From this system-level perspective, retracted citations serve as an indicator of how effectively signals of epistemic failure are surfaced, routed, and translated into reader-facing repair across human editors, automated tools, and platform infrastructure. While Wikipedia is often recognized for its agility in responding to breaking news and high-salience events \cite{keegan2012staying, twyman2017blm}, our findings reveal a much slower and uneven response to post-publication knowledge failures. Specifically, 71.6\% (845/1,181) of the retracted paper–page citation pairs were initially problematic—i.e., they were introduced before retraction or added after retraction without an in-text acknowledgment. By the end of the observation period, 32.9\% of these problematic citations remained uncorrected (278/845), while the rest were corrected through removal or in-text acknowledgment. Our survival shows that the median time to correction is 3.68 years, and a ``long tail'' of persistence indicates that (by the Kaplan–Meier estimate) 67.3\% of problematic citations remain uncorrected after approximately 1,000 days (2.74 years). These long delays reflect a gap between Wikipedia’s technical infrastructure for correction and the community's ability to mobilize editors. Given that Wikipedia continues to serve as a key access point for scientific information, this fragility in information repair raises essential concerns for both end users and the broader knowledge ecosystem.

Our findings can be understood through the lens of a classic coordination challenge in large-scale collaborative systems, where achieving quality is fundamentally dependent on successful coordination among contributors \cite{kittur2008harnessing}. This is more than a simple information routing issue, where public retraction notices fail to reach the right editors and articles \cite{ackerman2013sharing}. The challenge is also one of collective sensemaking and action \cite{keegan2013hot, weick2005organizing}. Even when editors are aware of a retraction, they may disagree on the appropriate response—whether to remove the citation entirely or add a warning \cite{viegas2004studying, kittur2008harnessing}—or on the urgency of the repair compared to other maintenance tasks, as volunteer attention is a scarce resource in large-scale peer production systems \cite{kittur2007he}. Accordingly, the long-tail persistence we observe is consistent with coordination frictions at multiple stages of repair: making the issue salient, agreeing on a reader-facing remedy, and allocating scarce attention to execute it. At the same time, our results suggest that these frictions are not uniform: when a retracted citation becomes more visible and actionable (e.g., through targeted signaling),  correction occurs substantially sooner, indicating that coordination can succeed once the problem is concretely surfaced and routed. Therefore, the persistence of these citations reveals a breakdown not only in information flow but also variability in the community’s ability to converge on a corrective strategy and prioritize action across pages and time. The remainder of this discussion interprets our empirical results through this framework, examining how different factors related to the paper, the retraction event, and the community are associated with this coordination.

Our analysis reveals that the prevalence of problematic retracted citations is not evenly distributed across academic domains. The Health Sciences domain exhibits the highest rate of retracted paper citations on Wikipedia relative to the total number of retracted papers in that field. In fact, retracted Health Science papers are over three times as likely to be cited on Wikipedia as non-retracted ones. This overrepresentation is particularly troubling given Wikipedia’s prominent role as a public health information source \cite{heilman2011Wikipedia}. Beyond prevalence, our Kaplan–Meier curves stratified by domain suggest that these domains also differ significantly in their correction time (see Figure \ref{delay_domain}). Citations in the Physical Sciences are corrected most rapidly, with a median persistence of approximately 228 days. In contrast, citations in the Life Sciences and Health Sciences display significantly longer persistence, with median correction times of 2,720 days and 1,846 days, respectively. The Social Sciences show intermediate behavior with a median of 724 days, though estimates in this domain should be interpreted cautiously given the smaller sample size. These results complicate a simple account of domain-specific collaborative maintenance. Specialized communities of practice, such as WikiProject Medicine \cite{heilman2011Wikipedia}, play an essential role in coordinating high-quality biomedical content. Yet, the Health Sciences still exhibit relatively slow correction rates. One plausible interpretation is that this accumulation reflects maintenance debt rather than systematic differences in source selection. Because retractions often occur long after papers are cited on Wikipedia, the ongoing task of monitoring previously accepted sources may exceed the capacity of manual editorial workflows alone. From this perspective, slower correction times in a domain with robust governance structures highlight the limits of relying on purely manual monitoring at scale and motivate the technical interventions discussed in Section \ref{sec:design}.

We also find that not all retractions are treated equally. In the extended Cox model, we do not observe a stable average difference in correction timing by salient retraction reasons at baseline. However, this association is time-dependent: the reason effect increases over the risk period, indicating that differences by retraction reason become more pronounced later rather than immediately after a citation enters the risk set. This suggests that citations to papers retracted for more salient reasons may be increasingly likely to be corrected as time passes, consistent with a delayed sensemaking process in which ``clearer'' epistemic failures become easier to justify and act upon after the retraction has had time to circulate. This editing pattern reflects a common trend in collaborative work during crises, in which people focus more on problems that seem urgent or clearly defined and are less likely to act when issues are more complex to recognize or explain \cite{twyman2017blm, keegan2013hot}.

A critical and counterintuitive finding is the limit of the ``wisdom of the crowd'' in this specific context. Contrary to the expectation that ``given enough eyeballs, all bugs are shallow,'' we found that the number of unique editors was not a significant predictor of correction speed. Simply having a larger volume of contributors does not accelerate the repair of deep bibliographic errors. Instead, we found that structural integration matters: citations on pages with a higher categories count are corrected significantly faster. This suggests that for maintenance tasks, organizational structure triumphs over raw participation volume \cite{kittur2008harnessing, halfaker2013rise}. Pages that are well-categorized are likely better integrated into Wikipedia’s ontology, making them more visible to experienced editors and WikiProjects who perform systematic cleanup \cite{forte2012coordination, kittur2008harnessing}. Furthermore, explicit signals of dispute proved highly effective. Citations flagged as controversial at the time of addition had a significantly higher hazard of correction. Unlike general editor volume, visual cues of disagreement successfully attract attention to specific bibliographic errors, surfacing them from the background noise of page history.

A key consideration is the role of automated tools, such as RetractionBot \cite{Headbomb2024Signpost}. This bot exemplifies the specialized roles that automated agents play in Wikipedia's quality control ecosystem \cite{zheng2019roles}. It primarily serves an editorial audience. By systematically adding a prominent template to the reference section, the bot is highly effective at its specific task of signaling a source's retracted status to editors engaged in verification and maintenance. The promise of such tools lies in this efficiency for well-defined tasks, and the community has become highly reliant on them for routine maintenance. For instance, the temporary failure of another specialized bot, an anti-vandalism tool, was shown to nearly double the time it took for the Wikipedia community to revert malicious edits \cite{geiger2013levee}. Similarly, Nyayachavadi et al. \cite{nyayachavadi2022characterizing} demonstrated the community's dependence on InternetArchiveBot to manage technical decay, such as `permanently dead' links. Our results powerfully validate this reliance in the context of epistemic repair. We found that citations flagged by RetractionBot had the highest hazard of correction, outperforming all other factors. This finding helps resolve the sociotechnical challenge of translating automated alerts into human judgment: it suggests that the ``technical flag'' applied by the bot serves as an effective scaffold for the substantive, contextual repair required to update the main text. Taken together, our findings suggest a two-stage process: automated monitoring helps surface candidates for review, and human/editorial work performs the reader-facing repair. The bottleneck may lie in coverage-which retracted citations are surfaced for attention—rather than in editors’ willingness to act once a credible signal is present.

Our most counterintuitive finding further underscores the necessity of better workflow design: 20.1\% of problematic citations were introduced after the paper had already been officially retracted. This phenomenon persists despite clear publisher-level signals, such as prominent ``RETRACTED'' watermarks on journal websites. We speculated that this disconnect stems from specific gaps in editorial workflows that allow editors to inadvertently bypass these warnings. For instance, editors may rely on locally stored PDFs downloaded prior to retraction, or engage in secondary citation by copying bibliographic details from other academic works without verifying the sources online. In these scenarios, the editor does not visit the publisher's landing page. While the failure to repair older citations might be attributed to passive under-maintenance, the active introduction of invalidated knowledge highlights a critical vulnerability: relying solely on external publisher signals is insufficient. This necessitates platform-level interventions detailed in Section \ref{sec:design}, such as embedding contextual alerts within the editing interface to detect these errors at the time of citation.

Our analysis also points to the role of external visibility cues in prompting correction. Citations to papers that received significant social media attention, particularly on Twitter/X, were associated with a significantly higher hazard of correction. This reflects the broader influence of public visibility on editorial prioritization: while social media activity can serve as a direct signal surfacing specific retractions \cite{zheng2025can}, it also acts as a proxy for a paper's general prominence. High-profile papers naturally attract greater scrutiny across the web, which may increase the probability that their retraction is noticed and addressed on Wikipedia. Crucially, our modeling reveals that this ``visibility advantage'' is ephemeral. We found a significant negative interaction with time, indicating that the strong initial association between tweet volume and correction hazard diminishes rapidly. In contrast, Open Access availability appears to provide a more sustained structural advantage. This aligns with recent scholarship by Yang et al. \cite{yang2024open}, who found that open access reduces barriers to access for volunteer editors. Furthermore, studies by Zheng et al. \cite{zheng2024gold, zheng2024comparative} demonstrate that Gold Open Access accelerates the ``academic purification'' process within the scientific community itself, suggesting that accessibility can facilitate earlier detection and “purification” upstream. Taken together, these evidences are consistent with a broader “transparency advantage”: accessibility may lower informational friction both in formal correction processes and in downstream venues like Wikipedia, although our observational design does not isolate a specific causal pathway. Conversely, not all forms of visibility facilitate repair: papers with higher academic citation counts associated with a lower hazard of correction.This is perhaps because their claims are more accepted, and editors are more hesitant to challenge them. These patterns emphasize the importance of salience: it is not just the existence of a retraction that matters, but its discoverability and the willingness of the community to act upon it.

Together, these findings highlight a multilayered vulnerability in Wikipedia’s citation infrastructure—not a simple story of individual neglect, but a coordination problem shaped by how epistemic risk signals become (or fail to become) discoverable and actionable. This vulnerability emerges from the interplay between visibility cues, structural organization (rather than sheer editor volume), bot-mediated flagging and maintenance workflows, and the salience of retraction events. Ultimately, this challenge is not unique to Wikipedia. Similar coordination issues likely exist in other distributed knowledge systems, such as open-source software projects managing bug fixes across dependencies, or collaborative fact-checking platforms striving to correct misinformation that has already propagated widely \cite{forte2012coordination, kittur2008harnessing}. Future work should continue to explore how to better surface and prioritize these risks, especially in domains where epistemic failure carries real-world consequences. Our work thus uses Wikipedia as a case study to illuminate a general vulnerability in the maintenance of large-scale, collaboratively produced knowledge, highlighting the critical need for systems that not only detect errors but also facilitate the complex social coordination required to repair them \cite{ackerman2013sharing, jackson201411}.

\subsection{Design Implications}
\label{sec:design}

Our findings suggest several design opportunities to address a fundamental coordination challenge: efficiently identifying and repairing citations to retracted scientific papers within Wikipedia's maintenance ecosystem. The core issue is not a lack of information—retraction notices are public and trackable—but a two-fold coordination challenge: first, a failure in connecting this critical information with the specific contexts and contributors responsible for repair; second, the subsequent challenge of achieving collective sensemaking to agree upon and prioritize a corrective action. Our survival analysis quantifies the consequences of this breakdown, revealing that correction delays can last for years. Crucially, our model demonstrates that signals of human attention may bridge the persistent gap between information availability and coordinated action. The following design directions are therefore aimed at strengthening the sociotechnical mechanisms that facilitate this crucial routing and coordination work.

\subsubsection{Improving Retraction Visibility in Context}
Our analysis reveals that early, human-generated warnings are effective; citations flagged as controversial are corrected significantly faster. However, such in-text flags are applied inconsistently. This suggests a need for tools that make these contextual warnings easier to create and more visible. To solve this information routing problem, Wikipedia’s editing interface could surface retraction indicators directly within citation markup or during editing interactions. 

Ideally, this functionality could be integrated into existing MediaWiki services. For instance, the ``Citoid'' service, which automatically retrieves citation metadata from URLs and DOIs, could be improved to check the retraction status of sources during metadata retrieval automatically. Furthermore, by leveraging the recent deployment of ``Edit Check,'' which prompts users to add citations for new claims, a similar ``Retraction Check'' mechanism could be implemented to verify the status of added sources in real time. By embedding retraction cues within the editing workflow, these changes would align with CSCW insights about situated action and the importance of interface-level nudges for distributed maintenance work \cite{geiger2017operationalizing}.

\subsubsection{Prioritizing Structure over Traffic in Maintenance Tools}

Our analysis challenges the assumption that high-traffic or highly edited pages are inherently self-correcting. We found that the number of unique editors was not a significant predictor of repair speed, suggesting that ``many eyes'' do not necessarily spot deep bibliographic errors. Instead, the significance of Categories Count indicates that structural integration is strongly asociated with maintenance. Pages that are well-categorized are likely more visible to WikiProjects and experienced ``gnome'' editors who use category-based tools to perform systematic cleanup.

This has direct implications for the design of community dashboards and task lists. Current tools often prioritize pages based on traffic or recent activity. Our findings suggest a different approach: maintenance tools should prioritize structural isolation. Algorithms could flag retracted citations on pages with low category counts (i.e., ``orphaned'' from the maintenance ontology) and high-risk features (e.g., Health Sciences domain, high citation count). By directing attention to these structurally disconnected pages—which our model shows are prone to neglect—dashboards can help extend coverage and route attention to structurally disconnected pages, where correction is otherwise likely to lag \cite{teblunthuis2018revisiting}.

\subsubsection{Automating In-Text Credibility Cues through Structured Metadata}
While editors can manually apply retraction templates, there is no consistent or automated mechanism to propagate retraction warnings across articles that cite the same paper. Our study reveals that corrections are uneven and often delayed, sometimes by years, resulting in gaps in credibility signaling. To reduce this burden, citation templates could be expanded to automatically ingest and display structured metadata from Crossref, Retraction Watch, or similar sources. If a DOI is retracted, the template could render a default warning, either inline or as a footnote, linking to the retraction notice. This would both standardize correction practices and reduce editors' cognitive load. One promising route is to leverage community-maintained structured sources within the Wikimedia ecosystem (e.g., Wikidata: WikiProject Retractions) as an intermediary layer for propagating retraction status across pages and interfaces.

\subsubsection{Surfacing External Signals for Epistemic Risk}
Finally, we find that external signals of visibility and accessibility are linked to correction speed. Retracted papers mentioned more frequently on Twitter/X are corrected significantly faster, while paywalled papers are corrected more slowly. This also suggests a time-sensitive design opportunity: external attention signals may be most useful as early-warning cues, and systems should surface them quickly before their association with repair fades. This pattern is consistent with the idea that social media attention may help surface retractions to editors, while a lack of open access may act as a barrier to verification and repair. Future tools might incorporate external alerting systems, integrating tweet activity and public discussions on platforms such as blogs or PubPeer regarding specific retractions into internal Wikipedia tools. This could help editors recognize emerging credibility issues even in articles that lack sustained editorial engagement.

Wikipedia’s infrastructure does not lack data about retractions—it lacks mechanisms to surface, route, and act on that data within existing workflows. By embedding signals into interfaces, prioritizing repair work through community-facing tools like dashboards and watchlists, and linking Wikipedia’s knowledge infrastructure with external retraction metadata, designers and editors can enhance the platform’s capacity for timely and collaborative epistemic repair.

\subsection{Limitations and Future Work}
While comprehensive in its analysis of the persistence and correction of retracted paper citations on Wikipedia, this study has several limitations that warrant further investigation.

First, our dataset is based on integrations across Retraction Watch, Crossref Event Data, Altmetric, and OpenAlex. While these sources provide broad coverage of retraction events and citation metadata, they are not exhaustive. In particular, both Crossref and Altmetric primarily track DOI-based citations, which means that retracted papers without DOIs, such as older publications or those in certain disciplines, are likely underrepresented in our analysis. Similarly, Wikipedia citations that refer to retracted papers through PubMed IDs, title strings, or bare URLs may be missed if they are not linked through standard identifiers. Future work could expand coverage by incorporating additional retraction databases (e.g., PubMed retraction tags), parsing free-text citation formats, or combining structured detection with manual validation to better capture edge cases.

Second, although our modeling identifies statistically detectable associations with correction speed, it cannot fully account for the editorial processes and sociotechnical dynamics that shape citation repair on Wikipedia. Decisions to add, retain, or remove citations are often entangled with disputes over content reliability, page notability, or conflicting interpretations of sourcing policies. Because we infer repair from revision traces, we observe outcomes more directly than the deliberative processes that produce them. Our analysis does not surface these nuanced deliberations, which may vary across topics, editor groups, or article histories. Qualitative research—such as interviews with active editors, analysis of talk page discussions, or observational studies of how citation templates are used—could provide deeper insight into how retraction cues are interpreted and acted upon in practice. More concretely, future work could examine how retraction signals travel through editors’ day-to-day workflows—especially for bot-mediated flags. For example, qualitative and trace-based mixed methods could investigate (i) how editors first notice RetractionBot templates (e.g., via watchlists, category maintenance, WikiProject backlogs, or recent-changes patrolling), (ii) how the presence of a template is translated into a specific repair action (complete removal versus in-text acknowledgment), and (iii) what points of friction or disagreement arise when deciding whether a retracted source can still be used for non-claim-supporting purposes (e.g., documenting controversy). More broadly, this line of inquiry could extend to upstream, structured sources of retraction metadata in the Wikimedia ecosystem. For instance, future work could examine how retraction status is curated and coordinated in Wikidata (e.g., through the Wikidata WikiProject on retracted research \cite{wikidata_wikiproject_retractions}), and whether—and through what technical pathways—such structured signals propagate into Wikipedia’s citation templates, bots, and editor-facing workflows. In addition, follow-up work could trace repeated introductions of the same retracted paper across multiple articles to distinguish independent sourcing from cross-page copying or “secondary citation” practices (e.g., copying bibliographic details from other Wikipedia pages, external summaries, or locally stored PDFs). Such analyses would help clarify why a non-trivial share of citations are introduced after retraction despite publisher-level warnings, and would better ground design interventions in the realities of editors’ verification routines.

Another limitation lies in the scope of our survival modeling. While our extended Cox model with start–stop intervals and time-varying terms is effective for estimating associations and accommodating non-proportional hazards, it does not capture potentially complex, non-linear interactions among predictors. For instance, visibility signals, page structure, and retraction characteristics may combine in ways that are not well represented by linear log-hazard specifications. Future work could explore interaction-focused modeling, flexible approaches (e.g., spline-based effects or more flexible statistical models), or mixed-methods designs to uncover richer mechanisms behind citation repair.

Finally, we must acknowledge a critical limitation in our classification of ``non-problematic'' citations. Our definition relies on human readers' cognitive capacity to interpret in-text warnings (e.g., ``This paper was retracted'') and appropriately discount subsequent claims. However, this assumption may not hold for Large Language Models (LLMs) that utilize Wikipedia as a foundational training corpus. Recent scholarship indicates that generative AI models may fail to contextually ``unlearn'' retracted findings even when textual flags are present \cite{thelwall2025does}. Consequently, citations that we classify as ``corrected'' for humans readers may still pose a persistent risk of propagating misinformation into the training data of downstream AI applications. 

By addressing these methodological limitations, future research can develop a more holistic understanding of the complex interplay between individual actions, community norms, and platform design in the ongoing process of collaborative knowledge maintenance.

\section{Conclusion}
This study examined the presence and persistence of retracted paper citations on Wikipedia, and modeled the timing of reader-facing correction events using survival analysis. Our findings reveal that while some citations are corrected promptly, many persist uncorrected for extended periods, with a median time to correction of 3.68 years (1,344 days). In our extended Cox model, a higher hazard of correction is associated with citations flagged by RetractionBot, citations framed as controversial at the time of addition, open access availability, and greater Twitter/X attention—although the Twitter/X association is time-dependent and attenuates over time. Conversely, a high academic citation count is associated with a slower correction time, which may reflect the difficulty of challenging sources perceived as highly authoritative. We do not observe a stable average difference by retraction reason in the baseline effect; instead, the association appears time-varying, suggesting that retraction characteristics may become more predictive later in the risk period rather than uniformly across time. These findings highlight a fundamental disconnect between the public availability of retraction data and the community’s collaborative workflows, which often fail to translate this information into the reader-facing edits required for a complete repair. Addressing this challenge requires moving beyond simple error flagging toward a tighter integration of information systems and human editorial processes. Design interventions that surface different types of signals to support editorial work—such as contextual signals embedded in editing interfaces, structured data signals within citation templates, and prioritized attention signals on community dashboards—could significantly improve responsiveness to invalidated science. More broadly, our results suggest that citation repair timing covaries with visibility cues and page-level organization, pointing to opportunities to strengthen retraction legibility and routing at the point of editing while supporting sustained cross-page maintenance work. By combining better information routing with sustained human oversight, Wikipedia can more effectively manage citation-level credibility failures and strengthen its role as a reliable public knowledge platform.

\section*{Acknowledgments}
We are deeply grateful to Aaron Shaw for insightful discussions and suggestions. We also thank the anonymous reviewers at ICWSM and IC2S2, as well as the research talk participants at IC2S2, for their constructive comments that helped improve this manuscript.

\bibliographystyle{plain}
\bibliography{cscw}

\appendix

\setcounter{table}{0}
\renewcommand{\thetable}{A\arabic{table}}

\section{Survival Analysis Model Selection Details}

\begin{table}[H]
\caption{First round test of the Proportional Hazards Assumption. The test is based on Schoenfeld residuals \cite{schoenfeld1982partial}. A significant p-value (p < 0.05) indicates that a variable violates the assumption.}
\centering
\begin{tabular}{l c c c}
\toprule
\textbf{Variable} & \boldmath$\chi^2$ & \textbf{df} & \textbf{p-value} \\
\midrule
RetractionBot & 1.505 & 1 & 0.220 \\
\textbf{Tweet Mention Count (at risk start)} & \textbf{10.768} & \textbf{1} & \textbf{0.001} \\
\textbf{Open Access} & \textbf{13.106} & \textbf{1} & \textbf{0.0003} \\
Team Size & 0.585 & 1 & 0.444 \\
Time to Retraction (days) & 1.070 & 1 & 0.301 \\
Pre-Retraction Editorial Event & 0.103 & 1 & 0.748 \\
\textbf{Retraction Reason (salient)} & \textbf{16.592} & \textbf{1} & \textbf{4.6e-05} \\
\textbf{Unique Editor Count (at risk start)} & \textbf{4.714} & \textbf{1} & \textbf{0.030} \\
Controversy Noted Initially & 0.346 & 1 & 0.556 \\
Media Count (at risk start) & 2.118 & 1 & 0.146 \\
\textbf{Categories Count (at risk start)} & \textbf{10.857} & \textbf{1} & \textbf{0.001} \\
\textbf{Reference Count (at risk start)} & \textbf{7.441} & \textbf{1} & \textbf{0.006} \\
Academic Citation Count (at risk start) & 0.732 & 1 & 0.392 \\
\textbf{Life Sciences (domain)} & \textbf{33.160} & \textbf{1} & \textbf{8.5e-09} \\
Physical Sciences (domain) & 1.706 & 1 & 0.191 \\
Social Sciences (domain) & 1.090 & 1 & 0.296 \\
\midrule
\textbf{GLOBAL} & \textbf{64.828} & \textbf{16} & \textbf{7.9e-08} \\
\bottomrule
\end{tabular}
\label{tab:ph_test_first}
\end{table}

\begin{table}[H]
\caption{Second round test of the Proportional Hazards Assumption. The test is based on Schoenfeld residuals \cite{schoenfeld1982partial}. A significant p-value (p < 0.05) indicates that a variable violates the assumption.}
\centering
\begin{tabular}{l c c c}
\toprule
\textbf{Variable} & \boldmath$\chi^2$ & \textbf{df} & \textbf{p-value} \\
\midrule
RetractionBot & 0.003 & 1 & 0.9577 \\
\textbf{Tweet Mention Count (at risk start)} & \textbf{10.755} & \textbf{1} & \textbf{0.0010} \\
\textbf{Open Access} & \textbf{5.197} & \textbf{1} & \textbf{0.0226} \\
Team Size & 0.245 & 1 & 0.6204 \\
Time to Retraction (days) & 0.299 & 1 & 0.5846 \\
Pre-Retraction Editorial Event & 0.315 & 1 & 0.5744 \\
\textbf{Retraction Reason (salient)} & \textbf{5.266} & \textbf{1} & \textbf{0.0218} \\
Unique Editor Count (at risk start) & 0.186 & 1 & 0.6659 \\
Controversy Noted Initially & 0.065 & 1 & 0.7981 \\
Media Count (at risk start) & 0.325 & 1 & 0.5687 \\
Categories Count (at risk start) & 2.240 & 1 & 0.1345 \\
Reference Count (at risk start) & 3.047 & 1 & 0.0809 \\
Academic Citation Count (at risk start) & 0.042 & 1 & 0.8384 \\
\midrule
\textbf{GLOBAL} & \textbf{28.147} & \textbf{13} & \textbf{0.0086} \\
\bottomrule
\end{tabular}
\label{tab:ph_test_second}
\end{table}

\section{Independent Variable Multicollinearity}

\begin{table}[H]
\centering
\caption{Collinearity diagnostics based on Lee \& Weissfeld (1996) for the time-dependent Cox model \cite{yul1996multicollinearity}. The analysis assesses the condition of the design matrix including the time-dependent covariate. The maximum Condition Index (CI) is 2.88, which is well below the strict threshold of 10 recommended for time-dependent models, confirming that multicollinearity is not a concern. CI diagnostics are based on model-based covariance (non-robust); robust SE are used for hypothesis tests and confidence intervals in the main model.}
\label{tab:ci_diagnostics}
\begin{tabular}{l c c}
\toprule
\textbf{Variable} & \textbf{Variance Proportion} & \textbf{Status} \\
& \textbf{(Max CI = 2.88)} & \\
\midrule
Academic Citation Count (at risk start) & 0.615 & Safe \\
Media Count (at risk start) & 0.235 & Safe \\
Tweet Mention Count (at risk start) & 0.103 & Safe \\
Controversy Noted at Initial Citation & 0.012 & Safe \\
Pre-Retraction Editorial Event & 0.008 & Safe \\
Open Access & 0.008 & Safe \\
Categories Count (at risk start) & 0.006 & Safe \\
Time to Retraction (days) & 0.004 & Safe \\
Reference Count (at risk start) & 0.003 & Safe \\
RetractionBot Flagged & 0.002 & Safe \\
Retraction Reason (salient) & 0.002 & Safe \\
Number of Unique Editors (at risk start) & 0.001 & Safe \\
Team Size & <0.001 & Safe \\
\bottomrule
\multicolumn{3}{l}{\footnotesize \textit{Note: Multicollinearity is indicated when two or more variables have variance proportions $>0.5$}} \\
\multicolumn{3}{l}{\footnotesize \textit{associated with the same large Condition Index ($>10$).}} \\
\end{tabular}
\label{tab:CI}
\end{table}

\section{Extended Cox Model Results}

\begin{table}[H]
\centering
\caption{Extended Cox proportional hazards model predicting time to correction (start–stop format). RetractionBot is modeled as a time-dependent covariate (0 pre-flag, 1 post-flag). Time-varying coefficients are modeled via covariate × log(time) interactions (denoted by ``X Time''). Robust (sandwich) standard errors clustered by citation pair are reported \cite{lin1989robust}.
}
\label{tab:results}
\begin{tabular}{l c c}
\toprule
\textbf{Variable} & \textbf{Hazard Ratio (95\% CI)} & \textbf{\textit{p}-value} \\
\midrule
\multicolumn{3}{l}{\textit{Paper Features}} \\
Tweet Mention Count & 1.62 (1.35 -- 1.94) & $< 0.001^{***}$ \\
\hspace{3mm} $\times$ Time & 0.97 (0.95 -- 1.00) & $0.032^{*}$ \\
Open Access & 1.68 (1.11 -- 2.55) & $0.015^{*}$ \\
\hspace{3mm} $\times$ Time & 1.01 (0.95 -- 1.07) & 0.778 \\
Team Size & 1.13 (0.99 -- 1.29) & 0.061 \\
Academic Citations (Prior Risk) & 0.85 (0.73 -- 0.97) & $0.020^{*}$ \\
\addlinespace

\multicolumn{3}{l}{\textit{Retraction Event Features}} \\
Time to Retraction & 1.16 (0.99 -- 1.37) & 0.063 \\
Pre-Retraction Editorial Event & 0.98 (0.69 -- 1.41) & 0.921 \\
Retraction Reason (Salient) & 0.97 (0.68 -- 1.40) & 0.881 \\
\hspace{3mm} $\times$ Time & 1.07 (1.01 -- 1.13) & $0.018^{*}$ \\
\addlinespace

\multicolumn{3}{l}{\textit{Wikipedia Page Features}} \\
RetractionBot Flagged (time-dependent covariate) & 2.34 (1.91 -- 2.86) & $< 0.001^{***}$ \\
Controversy Noted at Initial Citation & 1.63 (1.03 -- 2.60) & $0.039^{*}$ \\
Categories Count & 1.27 (1.11 -- 1.45) & $< 0.001^{***}$ \\
Reference Count & 1.05 (0.91 -- 1.20) & 0.534 \\
Unique Editors & 1.05 (0.92 -- 1.21) & 0.436 \\
Media Count & 0.97 (0.85 -- 1.10) & 0.614 \\
\bottomrule
\multicolumn{3}{l}{\footnotesize Significance codes: $^{***} p<0.001$, $^{**} p<0.01$, $^{*} p<0.05$} \\
\end{tabular}
\end{table}

\end{document}